\documentclass[aps,preprint,pre]{revtex4}
\usepackage{graphicx,color,epsfig}
\usepackage{amsfonts}
\usepackage{float}
\usepackage{amsmath}
\newcommand{\Ang}
   {\mbox{\AA}}

\begin{document}

\title[PRETRANSITIONAL BEHAVIOR IN A WATER-DDAB-5CB MICROEMULSION
CLOSE TO THE DEMIXING TRANSITION. EVIDENCE FOR INTERMICELLAR
ATTRACTION MEDIATED BY PARANEMATIC FLUCTUATIONS] {PRETRANSITIONAL
BEHAVIOR IN A WATER-DDAB-5CB MICROEMULSION CLOSE TO THE DEMIXING
TRANSITION. EVIDENCE FOR INTERMICELLAR ATTRACTION MEDIATED BY
PARANEMATIC FLUCTUATIONS}

\author{M.Caggioni$^1$, A. Giacometti$^2$, T. Bellini$^{1,3}$, N.A. Clark$^3$, F. Mantegazza$^4$, A. Maritan$^5$}

\address{$^1$Dipartimento di Chimica Biochimica e Biotecnologie per la Medicina, Universit\`{a} di Milano, via F.lli Cervi 93, 20090, Segrate (MI), Italy and INFM}

\address{$^2$Dipartimento di Chimica Fisica, Universit\`{a} di Venezia, S. Marta DD 2137, I-30123 Venezia, Italy and INFM}

\address{$^3$Department of Physics and Ferroelectric Material Research Center, University of Colorado, Boulder, Colorado 80309, USA}

\address{$^4$DIMESAB, Universit\`{a} di Milano–-Bicocca, 20052 Monza (MI), Italy and INFM}

\address{$^5$Dipartimento di Fisica "G. Galilei" dell'Universit\`{a} di Padova, via Marzolo 8, 35131 Padova, Italy and INFM}

\date{\today }

\begin{abstract}
We present a study of a water-in-oil microemulsion in which
surfactant coated water nanodroplets are dispersed in the
isotropic phase of the thermotropic liquid crystal 5CB. As the
temperature is lowered below the isotropic to nematic phase
transition of pure 5CB, the system displays a demixing transition
leading to a coexistence of a droplet rich isotropic phase with a
droplet poor nematic. The transition is anticipated, in the high T
side, by increasing pretransitional fluctuations in 5CB molecular
orientation and in the nanodroplet concentration. The observed
phase behavior supports the notion that the nanosized droplets,
while large enough for their statistical behavior to be probed via
light scattering, are also small enough to act as impurities,
disturbing the local orientational ordering of the liquid crystal
and thus experiencing pretransitional attractive interaction
mediated by paranematic fluctuations. The pretransitional
behavior, together with the topology of the phase diagram, can be
understood on the basis of a diluted Lebwohl-Lasher model which
describes the nanodroplets simply as holes in the liquid crystal.
\end{abstract}

\maketitle

\section{Introduction}\label{Introduction}
Liquid Crystals (LC), because of their soft spontaneous ordering,
have been found to produce interesting properties when used in the
production of micro- or nano-structured heterogenous materials
\cite{pre1}. Relevant examples are given by the large
electro-optic susceptibility of LC confined in a solid matrix
(e.g. PDLC \cite{pre44}), by the novel thermodynamic properties of
LC in random solid networks (e.g. LC in silica aerogels
\cite{pre2,pre3}), by the unprecedented interparticle forces
observed between LC suspended micron-sized colloids (e.g.
surfactant stabilized water emulsions \cite{pre4}). In these three
geometries (confinement, bicontinuous structures, LC as suspending
medium), the organization is surface dominated. Given the soft LC
ordering, forces arising at the interfaces effectively compete
with bulk forces, yielding a wide combination of simultaneous
effects such as wetting and pre-wetting, critical fluctuations,
topological defects, surface phase transitions and elastic
deformation.

Within this framework, a new intriguing possibility has been
recently proposed: materials combining the tendency to molecular
ordering of thermotropic LC, with the tendency for self
aggregation of amphiphilic molecules in solution. These systems
are simply obtained by replacing the oil phase of
surfactant-water-oil mixtures with a thermotropic LC. When the
molecular structures of the constituents are chosen in an
appropriate manner, these mixtures organize in ways still largely
unexplored.

After the pioneering work by Marcus \cite{pre5},
water-surfactant-LC mixtures have been extensively studied first
by Yamamoto and Tanaka \cite{pre6}, and later by us \cite{pre7}.
Yamamoto and Tanaka reported the "transparent nematic" phase in a
composite nematic liquid crystal (penthyl-cyanobiphenyl, 5CB),
double tailed ionic surfactant (didodecyl-dimethyl-ammonium
bromide, DDAB), and water system. The surfactant, which tends to
form interfaces concave toward the water at organic-water
interfaces, forms a water-in-oil type microemulsion of inverse
micelles in the 5CB host. Upon cooling, these mixtures were
reported to exhibit two phase transitions as indicated by double
peaked DSC scans: first to a new optically isotropic transparent
nematic phase and then to the coexistence of bulklike nematic (N)
and surfactant-rich isotropic (I) phases. According to the
interpretation, the new phase is actually a nematic phase in which
elasticity and radial boundary conditions on the micelle surface
combine to produce a nematic structure where the director is
distorted on the $10 nm$ length scale. This picture appears quite
similar to that offered by the behavior of micron-sized spheres in
nematic media, implying that micellar nano-inclusions, despite
their size, comparable to the LC molecules, can also be accounted
for using concepts typically suited for larger length scales, such
as use of the molecular director, continuous elastic distortions
and surface anchoring. However, the additional results provided by
our investigation on the same system \cite{pre7} contradict the
notion of a "transparent nematic" phase and support a different
interpretation: upon cooling from an isotropic phase resembling an
ordinary microemulsion of inverse micelles, the system undergoes a
demixing transition in which the micelle-poor phase develops long
range nematic order. This phase behavior, already observed
\cite{pre10} and predicted \cite{pre11} for molecular scale
contaminants such as polymers, can be understood on the basis of a
very simple model, in which micelles are considered as "holes" in
the LC. Experiments and theory thus combine to picture this system
- LC with nano-inclusions - in a way radically different from its
analog on the micron scale: nano-scale impurities act much more as
molecular contaminants - lowering the local degree of LC ordering
- than as microparticles - characterized by specific surface
anchoring of the surrounding LC. Moreover we observe, in agreement
with theoretical predictions, the presence of pretransitional
micellar density fluctuations anticipating, on the high
temperature side, the demixing. Such fluctuations can be
understood as resulting from intermicellar attraction, in turn
depending on the orientational paranematic fluctuation of the LC
matrix. Paranematic fluctuation mediated interactions have never
been observed before, although the presence of Casimir-style
interactions is generally expected. However, because of the first
order nature of the I-N transition of thermotropic LC, the
correlation lengths developed in the I phase extend only up to
$10-20 nm$, making pseudo-Casimir forces quite difficult to
observe.

In this article we present a new body of results on the same
microemulsion (hereafter referred to as $5DH \mu em$) described in
Ref. \cite{pre7}. New data obtained by Electric Birefringence
Spectroscopy (EBS) and by static and dynamic Light Scattering
(LS), (Sections \ref{Electric birefringence results} and
\ref{Light Scattering Results: Extracting Paranematic and})
coherently match the description of the system previously offered
\cite{pre7}, as explained in the discussion (Section
\ref{discussion}). In Section \ref{Theoretical model}, we present
a theoretical analysis of a simple spin model, which has been
extended in comparison with the one described in Ref. \cite{pre7}
in few respects, as further elaborated in Section \ref{Theoretical
model}. The relevance of the model in the interpretation of the
experiment is discussed in Section \ref{discussion}. In Section
\ref{discussion} we also compare the behavior of $5DH \mu em$ - in
which the inverse micelles can be viewed as providing annealed
disorder to the LC - with the behavior of LC incorporating
quenched disorder as described in a wide body of previous work.

\section{Experiments}\label{Experiments}
\subsection{Sample preparation}\label{Sample preparation}
The system studied in this paper is a three component mixture,
whose components are 5CB (penthyl-cyanobiphenyl, $M_w$ = 249.36)
from Merck, DDAB (didodecyl-dimethyl-ammonium bromide, $M_w$ =
462.65) from Sigma-Aldrich and de-ionized distilled water. Samples
were prepared by adding 5CB to a well homogenized mixture of DDAB
and water at room temperature. The samples were then hermetically
closed, kept at  $\sim 60^{\circ} C$ and stirred to ensure
homogeneity.

The system behavior,  and particularly the transition temperature,
was found to be very sensitive to the concentration of water and
thus reliable sample preparation required careful control of
evaporation and hydration of components and containers. No other
critical matter resulted emerged in the preparation procedure. At
$60^{\circ} C$, the system was indeed found to easily mix and
promptly homogenize, yielding a uniform transparent isotropic
state.

All the data presented in this article will refer to sample on the
dilution line characterized by weight ratio $m_{DDAB}/m_{H_2O}=9$.
We will indicate with $\phi$ the weight ratio $\phi
=\frac{m_D}{(m_{5CB}+m_D)}$, where $m_D=m_{DDAB}+m_{H_2O}$ is the
mass of the LC dispersed material. Being $\rho_{DDAB} \sim 870
kg/m^3$ and $\rho_{5CB} \sim 1022 kg/m^3$ we can calculate the
volume fraction of dispersed material $\phi_V \sim \phi \cdot
1.17$, and the overall density of the dispersed material $\rho_D
\sim 881 kg/m^3$.

As it will be shown below, dispersed water and surfactant
aggregate into inverted micelles having a water core, whose number
density $N_{mic}$ can be calculated once the volume of the single droplet
($v_{mic}$) is known.

We prepared different set of samples for the different
experimental investigations i.e. determination of phase diagram,
Kerr constant measurements and light scattering experiments. We
have found that different samples prepared with the same method
display phase separation at almost identical temperatures, which
confirms that the adopted sample preparation procedure is
reliable.

\subsection{Volumetric determination of the phase diagram}\label{Volumetric determination of the phase diagram}

\begin{figure}
\begin{center}
\includegraphics[width=0.8\textwidth]{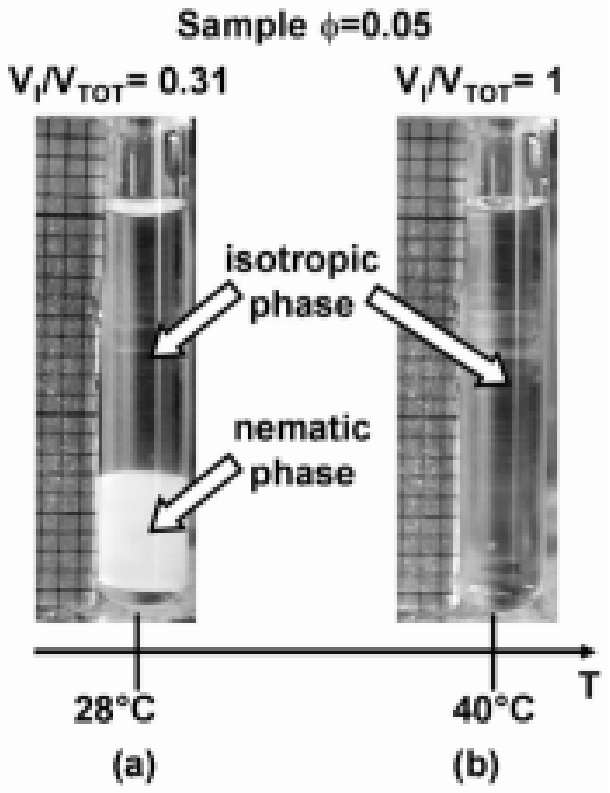}
\caption{\label{figura1} Pictures showing the phase separation of
$5DH \mu em$. At low temperature (a) the system consists of a
nematic phase filling 31 \% of the total volume, coexisting with a
lighter isotropic phase. At higher temperature (b) the system
appears as a single isotropic phase.}
\end{center}
\end{figure}

For temperature larger than the bulk $T_{NI}$, all the produced
$5DH \mu em$ samples are optically isotropic. As T is lowered, a
demixing phase transition takes place at a $\phi$ dependent
temperature $T_{DM}(\phi)$. At $T < T_{DM}(\phi)$ samples separate
into an isotropic phase coexisting with a denser optically
uniaxial (nematic) phase. By maintaining the samples at constant
temperature and waiting several hours, the two coexisting phases
macroscopically separate under the effect of gravity. The samples
appear as shown in Fig. \ref{figura1} (a), where the highly
scattering bottom nematic phase appears white. The two phases are
easily distinguishable and the relative volume can be measured by
taking pictures and analyzing the digital images. In order to
extract from volumetric measurements the phase boundary of the
coexistence region, we prepared - by successive dilution - four
different samples: $s_{1}$ $(\phi=0.02)$, $s_{2}$ $(\phi=0.05)$,
$s_{3}$ $(\phi=0.075)$, $s_{4}$ ($\phi=0.15$). These samples were
kept into a thermostatic chamber with a stability of $\pm 5 \cdot
10^{-3} K$ and observed by a digital camera at various
temperatures. In the inset of Fig. \ref{figura2} we show the T
dependence of the ratio, $V_I/V_{TOT}$, between the volume of the
isotropic phase and the total volume of the sample.
Sample concentrations were chosen so that, in a wide temperature
interval, at least two samples simultaneously have separated
phases large enough to be easily measured (typically larger than
$10 \%$ of the total volume). At each temperature we have
extracted the micellar weight ratios $\phi^I(T)$ and  $\phi^N(T)$,
enclosing the coexistence region of the $5DH \mu em$, shown
as dots in Fig. \ref{figura2}.

In this analysis we have assumed that the behavior of
$5DH \mu em$ can be understood as a pseudo-single component
system, i.e. that micellar aggregates don't change their structure
as the phase transition takes place, but simply redistribute
between the two phases. This assumption is confirmed by light
scattering measurements in the isotropic phase, as discussed in
the next Sections. Accordingly, from the standpoint of micelles,
the phase transition is a simple demixing in which they
preferentially accumulate in the isotropic phase. Since micelles
are less dense than 5CB, their presence increases the density
difference between isotropic and nematic phase.

\subsection{Electric Birefringence Spectroscopy}\label{Electric Birefringence Spectroscopy}
\begin{figure}
\begin{center}
\includegraphics[width=0.8\textwidth]{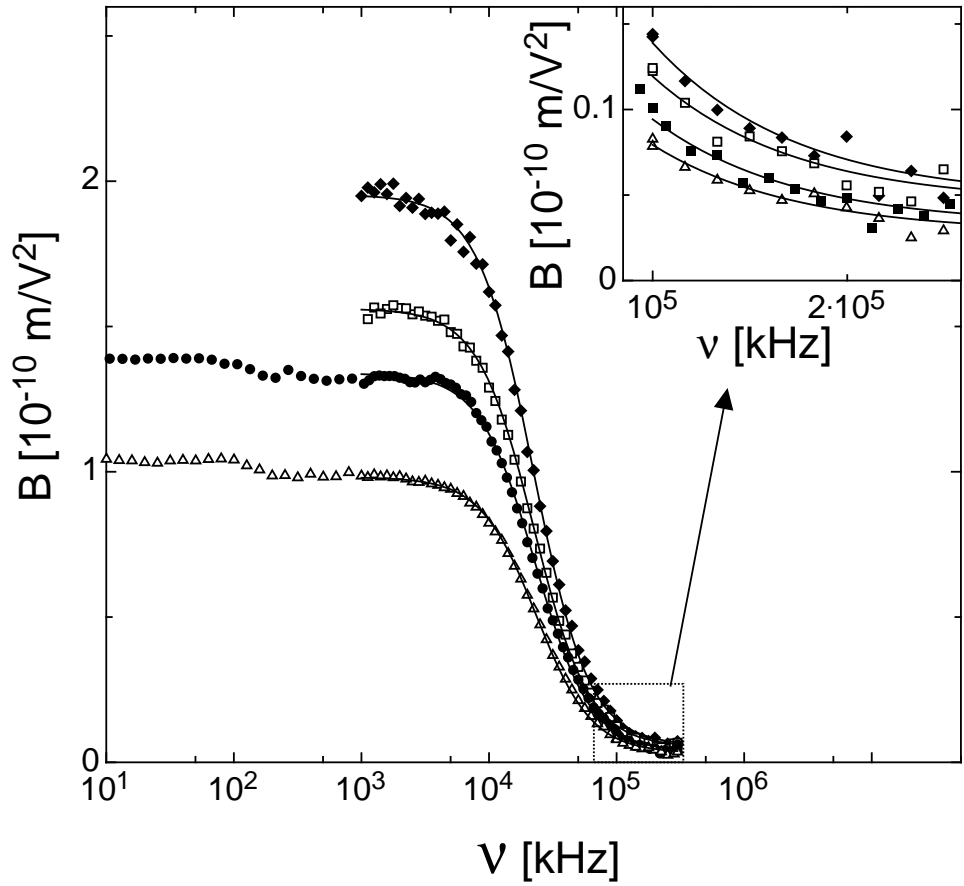}
\caption{Electric birefringence spectra measured on a sample of
bulk 5CB at various temperatures: $T=35.93^{\circ}C$ (triangles),
$T=35.53^{\circ}C$ (dots), $T=35.28^{\circ}C$ (squares),
$T=35.11^{\circ}C$ (diamonds). Lines: fittings with the Debye Eq.
\ref{debyekerr}. Inset: enlargement of the high frequency data and
fits. \label{figura3}}
\end{center}
\end{figure}

\begin{figure}
\begin{center}
\includegraphics[width=0.8\textwidth]{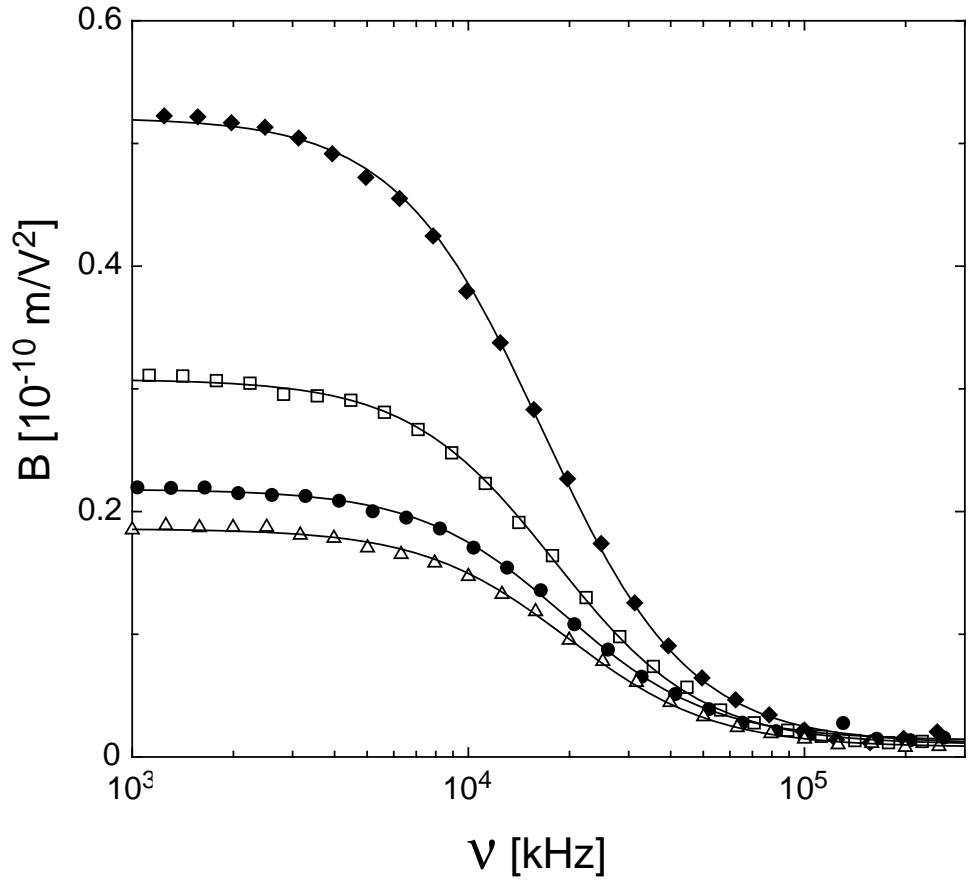}
\caption{\label{figura4}Electric birefringence spectra measured on
a sample of $5DH \mu em$ with $\phi=0.15$ at various temperatures:
$T=33.66^{\circ}C$ (triangles), $T=32.86^{\circ}C$ (dots),
$T=31.85^{\circ}C$ (squares), $T=30.80^{\circ}C$ (diamonds).
Lines: fittings with the Debye Eq. \ref{debyekerr}.}
\end{center}
\end{figure}

Measurement of the Kerr coefficient have been performed by
applying pulses of ac voltage. The pulse duration ($0.1 -– 10 ms$)
was abundantly enough for the system to reach the steady state.
The frequency $\nu$ of the applied field ranges from 1 kHz to 300
MHz, the upper limit being set by inductive loops introducing
distortions in the detection of the field applied to the cell
electrodes. For this reason the wiring has been designed to keep
wire loops to a minimum. The field induced birefringence is, at
steady state, composed by a dc component and by an ac component of
frequency $2 \nu$. While the former is related solely to the
amplitude of the local dielectric anisotropy characterizing the
nematic fluctuations in the isotropic phase, the latter is instead
strongly dominated by the dynamics of the paranematic
fluctuations. The set up used for these measurements is not
equipped with a fast enough detection optics and transient
digitizer to reliably measure the amplitude of the ac component.
It is instead very accurate in the measurement of the dc
component: the use of a quarter-wave plate in between cell and
analyzer, the automated set-up enabling measurements at various
angular positions of the analyzer and various field amplitudes, as
well as the accumulation of a large number of pulse response at
each studied frequency, allows extracting the field induced non
oscillatory component of the birefringence $\Delta n_{dc}
=n_{\parallel}-n_{\perp}$ with an error of few percent. The
measured $\Delta n_{dc}$, has been found, as expected, to be
proportional to the square of the electric field amplitude E,
hence enabling a meaningful extraction of the Kerr constant
$B(\nu)$ defined as:

\begin{equation}
B(\nu)=\frac{\Delta n_{dc}}{\lambda \langle E^2 \rangle}
\end{equation}

where   $\lambda = 633 nm$ is the optical wavelength used in the
experiment. In Figs. \ref{figura3} and \ref{figura4} we show
$B(\nu)$ spectra measured in bulk 5CB and in a $5DH\mu em$ sample.
As clearly visible, our experimental frequency range includes the
main features of the spectra.

\subsection{Static and Dynamic Light Scattering}\label{Static and Dynamic Light Scattering}
Light scattering measurements have been performed by a highly
automated experimental set up, featuring a He-Ne laser source, a
temperature control with a $5 \cdot 10^{-3} K$ stability and
collection optics coupling scattered light into optical fibers. We
used multimode fibers to measure the intensity of both polarized
and depolarized scattered light ($I_{VV}$ and $I_{VH}$
respectively). Single mode fibers were instead used to measure
intensity autocorrelation functions of the polarized and
depolarized scattered light ($G_{2,VV}$ and $G_{2,VH}$,
respectively). Use of fiber beam-splitter and cross correlation
has enabled to reliably extract correlation functions down to a
retardation time of 25 ns, a limit set by the electronics of the
BI9000 correlator board used in the experiment. All static and
dynamic measurements were performed at fixed angle $\theta =
90^{\circ}$.

\begin{figure}
\begin{center}
\includegraphics[width=0.8\textwidth]{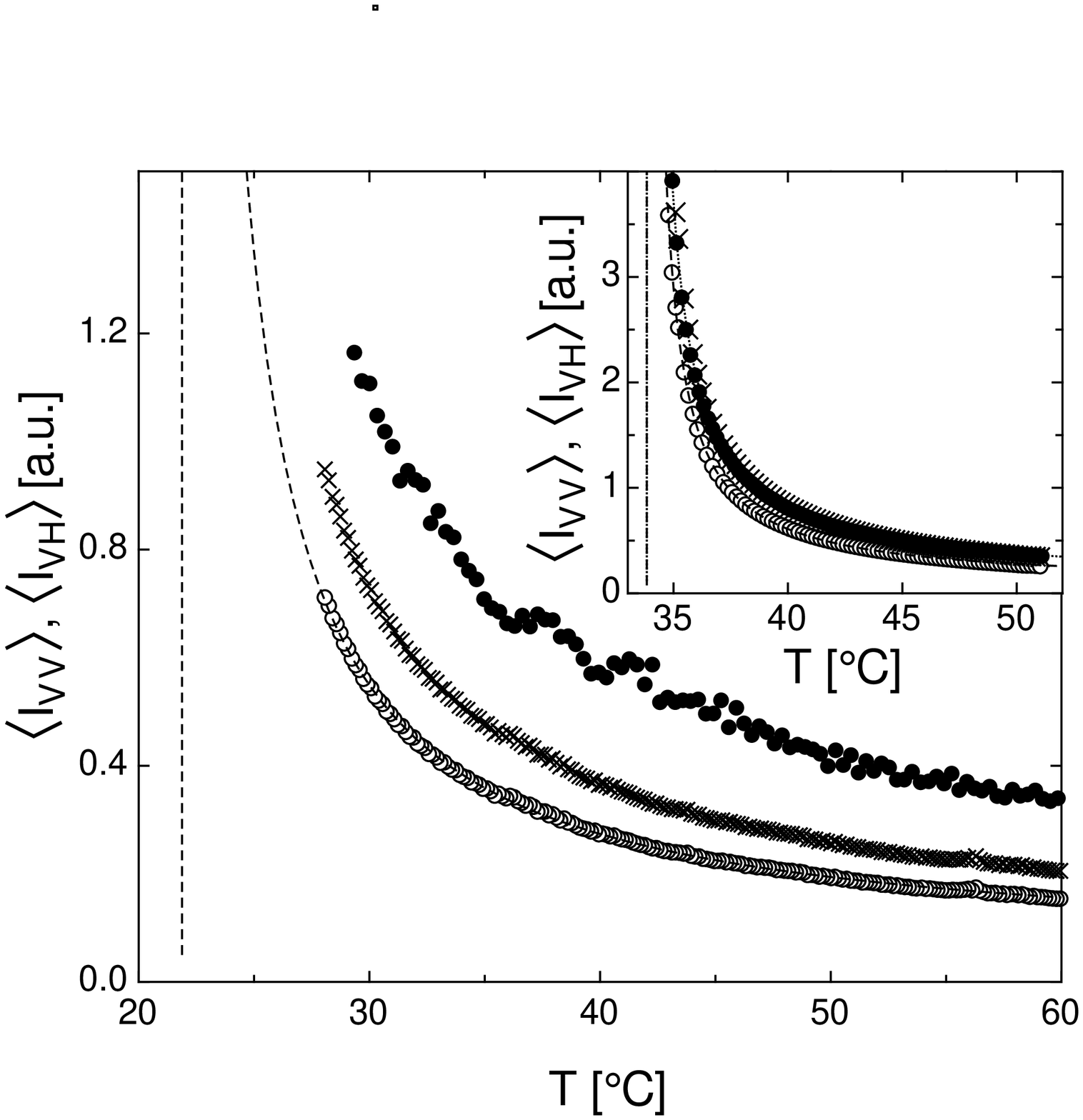}
\caption{\label{figura5} Scattered intensity in in a $5DH \mu em$
sample with $\phi=0.15$ (main figure) and in bulk 5CB (inset) as a
function of temperature. Empty dots: depolarized scattered
intensity $\langle I_{VH} \rangle$. Full dots: polarized scattered
intensity $\langle I_{VV} \rangle$. Crosses: depolarized scattered
intensity scaled by a $4/3$ factor (see Eq. \ref{ivvivh}). Lines:
fittings with $(T-T^*)^{-1}$. Main figure: $T^*=21.9^{\circ}C$,
indicated by the vertical line. Inset: $T^*=33.8^{\circ}C$, for
both the polarized and depolarized components.}
\end{center}
\end{figure}

\begin{figure}
\begin{center}
\includegraphics[width=0.8\textwidth]{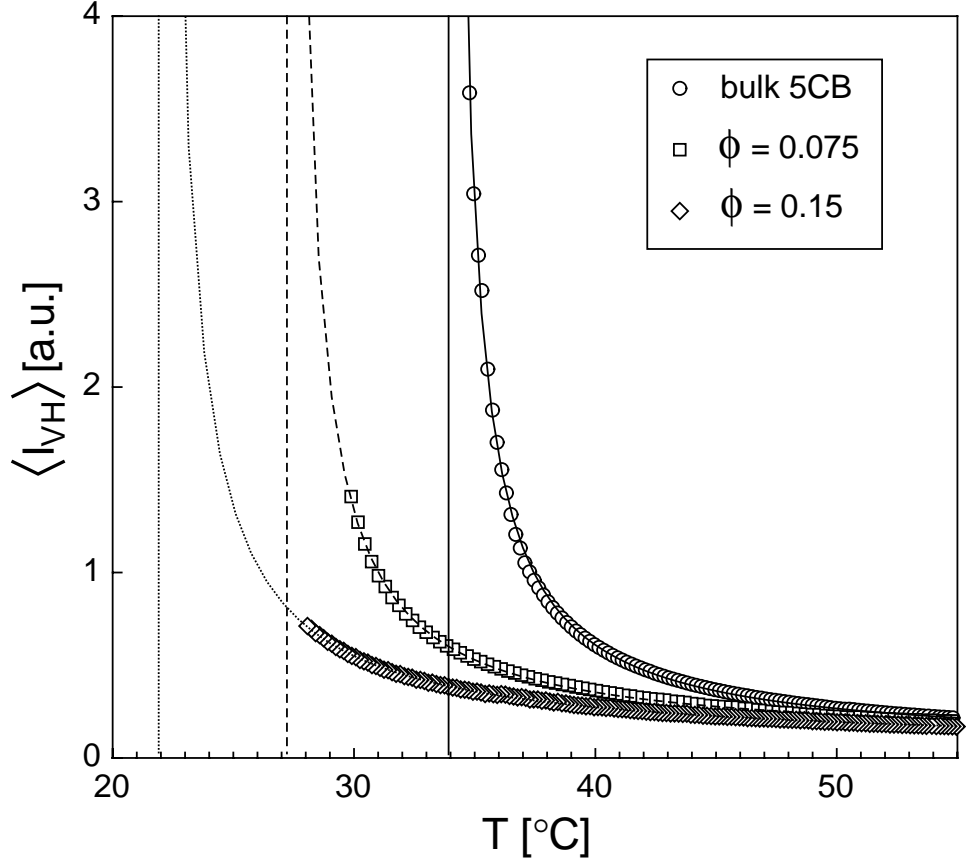}
\caption{\label{figura6}Comparison of the depolarized intensity
scattered by bulk 5CB (circles) and $5DH \mu em$ with $\phi=0.075$
(squares) and $\phi=0.15$ (diamonds) as a function of temperature.
Lines: fittings with $(T-T^*)^{-1}$ yielding
$T^*(bulk)=33.8^{\circ}C$ (solid vertical line),
$T^*(\phi=0.075)=27.2^{\circ}C$ (dashed vertical line) and
$T^*(\phi=0.15)=21.9^{\circ}C$ (dotted vertical line).}
\end{center}
\end{figure}

In Figs. \ref{figura5},\ref{figura6} we show the polarized and
depolarized scattered intensity for bulk 5CB and $5DH \mu em$
samples. While $I_{VV}$ and $I_{VH}$ measured in each sample at
various T are on the same intensity scale, comparison between
intensities scattered by different samples should be considered
only as semi-quantitative since cell replacement required
realignment of the collection optics.

\begin{figure}
\begin{center}
\includegraphics[width=0.8\textwidth]{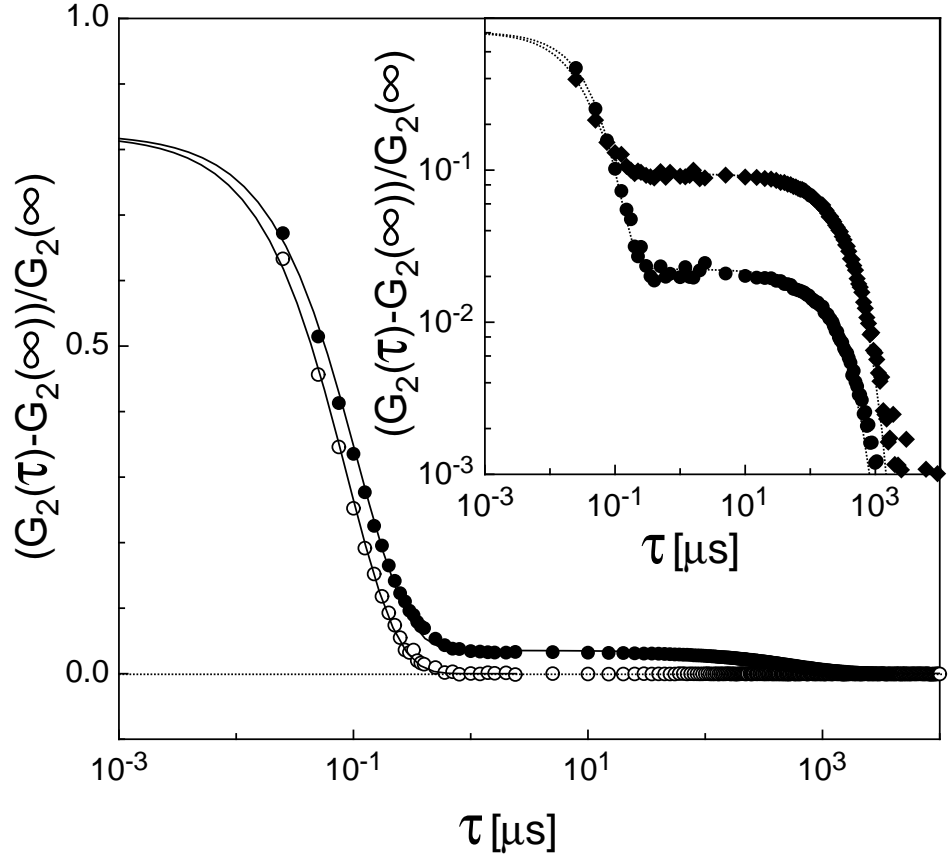}
\caption{\label{figura7} Normalized autocorrelation functions of
the polarized (VV, full dots) and depolarized (VH, open dots)
scattered intensities for the $\phi=0.075$ $5DH \mu em$ sample at
$T=31^{\circ}C$. Inset: comparison of normalized autocorrelation
functions of the polarized scattered intensity for the
$\phi=0.075$ (full dots) and $\phi=0.15$ (diamonds) $5DH \mu em$
samples at $T=31^{\circ}C$. Lines: fit with double exponential
decay (VV) and single exponential decay (VH) }
\end{center}
\end{figure}

Fig. \ref{figura7} shows typical $G_{2,VV}$ and $G_{2,VH}$
measured by studying bulk $5DH \mu em$ samples. The inset of Fig.
\ref{figura7} instead compares $G_{2,VV}$ as measured in $ \phi =
0.075$ and in $\phi = 0.15$ $5DH \mu em$.

\section{Description and analysis of the results}\label{Description and analysis of the results}

\subsection{The phase diagram}\label{The phase diagram}
\begin{figure}
\begin{center}
\includegraphics[width=0.8\textwidth]{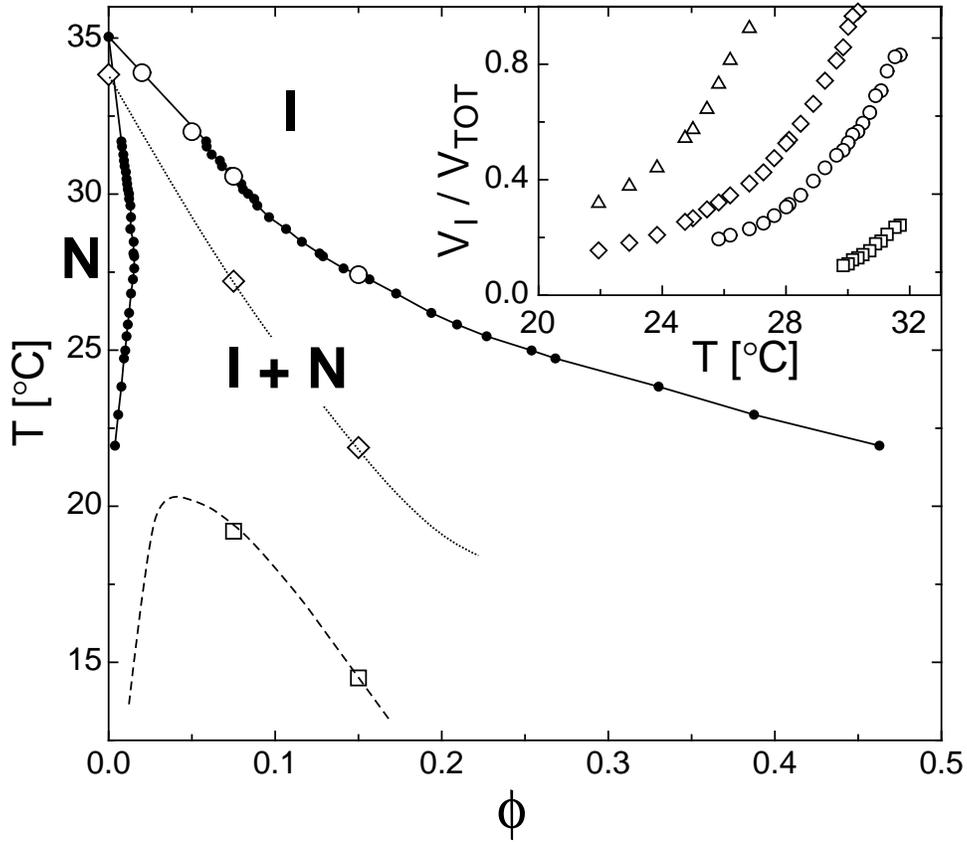}
\caption{\label{figura2}Experimental phase diagram featuring
isotropic (I), nematic (N, grey shadowing) and phase coexistence
(I+N). Full dots: phase boundaries obtained from the volumetric
measurements shown in the inset. Open dots: $T_{DM}$ of the four
samples used for volumetric measurements. Open diamond:
extrapolated divergence temperature $T^*(\phi)$ of paranematic
fluctuations. Open squares: extrapolated divergence temperature
$T_S(\phi)$ for micellar density fluctuations. Inset: volume
fraction of the isotropic phase in the two phase coexistence
region for: $\phi=0.02$ (squares), $\phi=0.05$ (circles),
$\phi=0.075$ (diamonds) and $\phi=0.15$ (triangles).}
\end{center}
\end{figure}

The most relevant features of the phase diagram in Fig. \ref{figura2} are the
coexistence of a micelle poor nematic and a micelle rich isotropic
over an extended temperature range and the presence of a
re-entrant nematic phase.

It is known that inclusion of molecular dopants in a nematogenic
compound results in extending the isotropic-nematic phase
coexistence - confined to $T = T_{NI}$ in bulk LC - to a wide T
interval. This phenomenon has been experimentally studied in
detail for low contaminant dilution of aliphatic molecules having
various shape \cite{pre8}. The authors also present a
thermodynamic analysis of the $T_{NI}(\phi)$ slopes of the
isotropic and nematic phase boundaries (approximated by straight
lines converging to $T_{NI}$). In particular they showed that
nematic-isotropic thermodynamic equilibrium implies that

\begin{equation}
\frac{1}{\beta_N}-\frac{1}{\beta_I}=\frac{\Delta S_{NI}}{R}
\label{eqmartire}
\end{equation}

where $\beta_{N,I}=\frac{dT^R_{N,I}}{dx}$ are the are the slopes
of the (respectively nematic and isotropic) phase boundaries
enclosing the coexistence region, $T^R=\frac{T}{T_{NI}}$, $x$ is
the contaminant concentration expressed in mole fraction, R is the
gas constant and  $\Delta S_{NI}$ is the N-I transition entropy
for bulk 5CB that can be extracted from calorimetric measurements,
namely $\Delta S_{NI}/R \sim 0.25$ \cite{pre8}.

According to Eq. \ref{eqmartire}, the quantity ($\beta_N^{-1} -
\beta_I^{-1}$) is independent from the nature of the contaminant.
Ref. \cite{pre8} reports that, for molecules of elongated shape,
$\beta_N \sim -0.58$ and $\beta_I \sim -0.51$, thus approximately
confirming the predicted behavior. A successive study of the
effect on the I-N transition of 5CB doped by surfactant molecules
of the polyoxethylene family reports $\beta_N \sim -1.35$ and
$\beta_I \sim -0.95$ \cite{pre9}, also is agreement with Eq.
\ref{eqmartire}.

The extrapolated initial slopes of the phase diagram in Fig.
\ref{figura2} are: $dT^R_N/d\phi \sim - 1.23$ and $dT^R_I/d\phi
\sim - 0.20$. By assuming that all contaminants have the molecular
weight of DDAB (i.e. disregarding the presence of water), we
obtain $\beta_N \sim -2.6$ and $\beta_I \sim -0.35$, while, if we
calculate exactly the total molar fraction of contaminants (water
included) we obtain $\beta_N \sim -0.75$ and $\beta_I \sim -0.10$.
In both cases the prediction of Eq. \ref{eqmartire} is not
confirmed since ($\beta_N^{-1} - \beta_I^{-1}$) is 10 to 30 times
larger than predicted. Indeed, the $5DH \mu em$ phase diagram is
evidently characterized by a separation between the isotropic and
the nematic phase boundaries much larger than those found in refs
\cite{pre8} and \cite{pre9}. This analysis suggests that, in
calculating $\beta$, we are overestimating the molar concentration
of dopants. Indeed, completely different estimates are obtained if
one considers, as contaminants, not the single monomeric dispersed
surfactant molecules, but the micellar aggregates, as discussed in
section \ref{discussion} after micellar sizes and interactions
are introduced.

Although predicted by many previous theoretical analysis, the
presence of a re-entrant nematic phase has been experimentally
observed only quite recently \cite{pre10} by studying mixtures of
the liquid crystal E7 and silicone oil. The nematic phase has been
found to extend to a maximum contaminant concentration of 3 \%
vol. at about $10^{\circ} C$ below $T_{NI}$, i.e. similar but
slightly larger than the one in Fig. \ref{figura2}. The physical
origin of reentrant nematic phase can be sought in the combination
of the positive T vs. $\phi$ nematic phase boundary expected at
the lowest temperatures with the negative T vs. $\phi$ phase
boundary resulting predicted for temperatures close to the NI
transition. While at the limit of low temperature the nematic
crystal expels all impurities, as T is increased the entropic gain
for impurity solubilization determines an increment of the maximum
micellar concentration in the N phase. Upon increasing further the
temperature, the free energy difference between the bulk N and I
phase is reduced, vanishing at $T_{NI}$. Therefore for each
density of impurities solubilized in the nematic at low T, a
maximum T is found where the free energy of the nematic with
impurities equals the energy of the isotropic phase. At this
temperature it pays for the system to phase separate into two
phases as described by equation \ref{eqmartire}. The cross-over
between the entropically driven impurity solubilization at low T
and the impurity driven phase instability at high T leads to the
observed reentrant behavior.

Phase diagrams similar to those in Fig. \ref{figura2} have also
been found in a somewhat different research contexts. The
contaminant induced I-N phase separation is exploited to produce
Polymer Dispersed Liquid Crystal materials. An isotropic mixture
of isotropic LC and polymers is destabilized by a T quench. The
isotropic phase hosting nematic nuclei is subsequently hardened by
photo or chemically induced cross polymer linking. Such a phase
separation has therefore received particular attention. Its
experimental characterization, performed by studying a mixture of
E7 and polybenzylmethacrylate \cite{pre11}, is qualitatively
similar to the isotropic phase boundary in Fig. \ref{figura2}
although few experimental points may suggest the presence of an
additional isotropic-isotropic phase separation.

\subsection{Electric birefringence results}\label{Electric birefringence results}

The onset of local nematic structuring, as the one predicted by
Yamamoto and Tanaka's "transparent nematic", would quite likely
result in an electro-optic susceptibility larger than the
conventional isotropic phase, as also suggested by the reported
shear birefringence of such a phase. This is why measurements of
electro-optic response by EBS appear particularly appropriate to
study the $5DW \mu em$.

\subsubsection{Bulk 5CB}\label{Bulk 5CB EBS}

According to the mean field description, the Kerr coefficient for
bulk LC is given by \cite{pre12}:

\begin{equation}
B(\nu) = \frac{\epsilon_0 \Delta \epsilon_r^X(\nu) \Delta n}{3
\lambda a (T-T^*)} \label{eqkerrlc}
\end{equation}

where $a$, $\Delta \epsilon_r^X$ and $\Delta n$ are, respectively,
the first coefficient in the Landau-DeGennes expansion and the
dielectric anisotropy and birefringence of a completely ordered
nematic. $\Delta \epsilon_r^X = \epsilon_{\parallel}^X -
\epsilon_{\perp}^X$, the parallel and perpendicular component of
the dielectric coefficient of a perfectly ordered nematic (the
superscript $X$ refer to nematic crystal), gauges the amplitude of
the coupling of E with the induced nematic order, and is the only
term in Eq. \ref{eqkerrlc}  depending on the frequency of the
applied electric field. However, the contribution of
$\epsilon^{X}_{\perp}$ to the frequency dependence of B is
negligible: the permanent dipole of 5CB is aligned with the
molecule and thus only electronic polarization contributes to
$\epsilon^X_{\perp}$, which is therefore not expected to have
observable frequency dependence in our experimental EBS spectra.
The dominant contribution of $\epsilon_{\parallel}^X$ arises from
molecular reorientation, its characteristic frequency deriving
from the kinetics of molecular reorientation around the short
axis. We thus expect EBS spectra for isotropic bulk 5CB to be
close to its dielectric spectra.

In short, Eq. \ref{eqkerrlc} indicates that EBS spectra enable
accessing two independent quantities: the dynamics of the
individual molecules from the frequency dependence of $B(\nu)$,
and the degree of local paranematic ordering from the low
frequency value of B. It is worth noticing that this last quantity
cannot be detected by ordinary dielectric spectroscopy.

With respect to the more conventional pulsed electric
birefringence, EBS better enables a distinction between the
various processes taking place in the different frequency ranges.
This is useful when low frequency artifacts, such as electrode
polarization and space charge accumulations, typically frequency
dependent, are present. In practice, our data, shown in Fig.
\ref{figura3}, show little frequency dependence in the kHz regime.
We have fitted, for $\nu > 1 $MHz the measured $B(\nu)$ with the
sum of a frequency independent value, $B_e$, expressing the Kerr
coefficient due to the anisotropy in the electronic polarization
of 5CB molecules, and of a Debye type relaxation describing the
frequency dependence of the polarization due to the reorientation
of permanent dipoles:

\begin{equation}
B(\nu)=B_e+\frac{B_d}{1+(2 \pi \frac{\nu}{\nu_c})^2}
\label{debyekerr}
\end{equation}

In the fitting procedure, $B_e$, $B_d$ and $\nu_c$ are used as
fitting parameters. The good quality of the fits indicates that a
single relaxation frequency is enough to describe the whole
polarization process. This may appear surprising since previous
investigations by high frequency and time-domain dielectric
spectroscopy on cianobiphenyls in the isotropic phase showed
evident deviations from Debye behavior \cite{pre13} \cite{pre14}
\cite{pre15}. In fact, Debye behavior is instead regularly
observed when studying the $\epsilon_{\parallel}$ spectra in the
nematic phase. We argue that, since $B(\nu)$ reflects the
frequency dependence of $\epsilon_{\parallel}^X$, it is not
actually unexpected that it behaves more similarly to the nematic
$\epsilon_{\parallel}$ than to the isotropic $\epsilon$. This
phenomenology together with more data on cianobiphenyls compounds,
will be discussed in a forthcoming publication \cite{pre16}.

The three fitting parameters are found to depend on temperature.
At all temperatures, we find that $B_e \sim B_d/30$. This ratio is
in rough agreement with what expected by comparing dielectric
$(\nu < \nu_c)$ and optical $(\nu > \nu_c)$ anisotropies: from
literature values \cite{pre44}, in the nematic phase
$(n_{\parallel}^2 - n_{\perp}^2)/(\epsilon_{\parallel} -
\epsilon_{\perp}) \sim 1/20$. The factor 1.5 in between estimate
is probably due to the fact that our experiments are performed in
the I phase, at values of field induce nematic order parameter
much smaller than achievable in the N phase $(S(E = 10 V/mm) \sim
1 \cdot 10^{-6})$ where the above ratio has instead been
evaluated. The temperature dependence of $B_e$ and $B_d$ is shown
in Fig. \ref{figura8}. As indicated, the uncertainty on $B_e$ is
much larger than for $B_d$, and its T dependence has to be
considered only qualitatively. In the same figure we have plotted
the fit of $B_d(T)$ to $A_{5CB}(T-T^*)^{-1}$, yielding $A_{5CB} =
1.9 10^{-10} mV^{-2}$ and $T^* = 33.4 \pm 0.1^{\circ} C$. Both the
quality of the fit and the value of $T^*$ in agreement to previous
determinations, confirms that the amplitude of the Kerr effect is
well described by a simple mean field theory.

In Fig. \ref{figura9} we plot  $\nu_c$ as a function of $1/T$. The
Arrhenius-like behavior,

\begin{equation}
\nu_c \propto e^{(-\Delta H / RT )}
\end{equation}

is evident, where  $\Delta H$ is the activation energy. We find
$\Delta H = 41.0 \quad kJ/mol$. This number should be compared
with $\Delta H = 33 \quad kJ/mol$, obtained by fitting the
Arrhenius dependence of the cut-off frequency of the 5CB
dielectric spectra in the isotropic phase \cite{pre45}. According
to mean-field expression, both EBS and dielectric characteristic
frequencies should be proportional to the fluid viscosity
\cite{pre55}. Such proportionality is partially confirmed when
$\Delta H_{\epsilon}$ is compared with activation energy extracted
from by the Arrhenius T dependence of 5CB viscosity: from
literature we get:  $\Delta H_{\eta} = 29.0 \quad kJ/mol$
\cite{pre19},  $\Delta H_{\eta} = 44.9 \quad kJ/mol$ \cite{pre20},
$\Delta H_{\eta} = 28.3 \quad kJ/mol$ \cite{pre21}. Despite the
rather surprising numerical discrepancies, it is very clear from
the data that EBS, in analogy to viscosity, probes the dynamical
response of individual molecules, undergoing activated processes
of angular reorientation. We will exploit the access to local
molecular dynamics through EBS to evaluate how much the micellar
inclusions affect the local viscosity, a notion necessary to
interpret micellar diffusion.

\subsubsection{$5DH \mu em$}\label{me EBS}

\begin{figure}
\begin{center}
\includegraphics[width=0.8\textwidth]{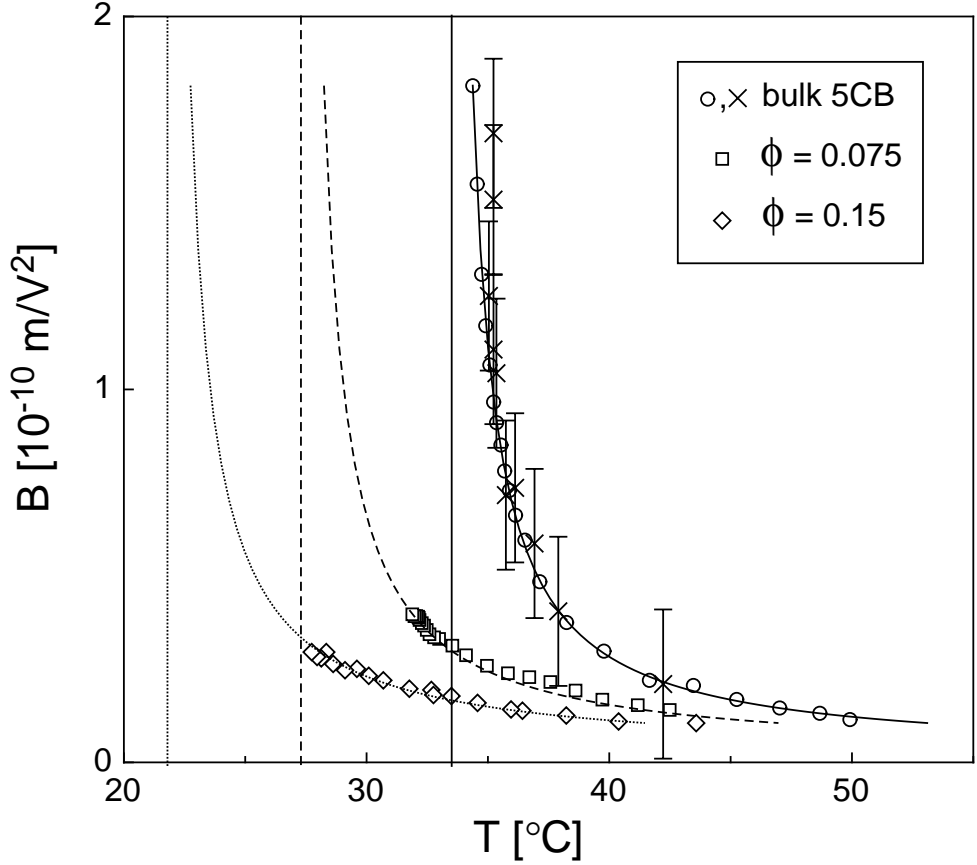}
\caption{Low frequency Kerr constant $B$ of Bulk 5CB (circles) and
$5DH \mu em$ with $\phi=0.075$ (squares) and $\phi=0.15$
(diamonds) as a function of temperature. Crosses: high frequency
Kerr constant asymptote $B_e$ of bulk 5CB multiplied by a factor
15 to overlap $B_d$. Lines: fittings with $(T-T^*)^{-1}$, as
detailed in the text, yielding $T^*(bulk)=33.4^{\circ}C$ (solid
vertical line), $T^*(\phi=0.075)=27.3^{\circ}C$ (dashed vertical
line) and $T^*(\phi=0.15)=21.8^{\circ}C$ (dotted vertical line).
\label{figura8}}
\end{center}
\end{figure}

As a simple inspection of the data in Fig. \ref{figura8}
immediately reveals, EBS results for the DDAB, water and 5CB
mixtures are qualitatively similar to the bulk data although
depressed in temperature and limited in amplitude. Although this
is already enough to make clear that no extravagant molecular
organization is present in the system, it is nevertheless
interesting to analyze the results as it has been done for bulk
5CB.

All EBS spectra can be fitted with Debye dispersions (Eq.
\ref{debyekerr}) as in the bulk. In Fig. \ref{figura8} we show the
low frequency Kerr coefficient $B_d$ as a function of temperature.
The lowest value achieved for each given $\phi$ ($\phi = 0.15$ and
$\phi = 0.075$) corresponds to the temperature at which demixing
occurs, $T_{DM}(\phi = 0.075) \sim 31^{\circ} C$ and $T_{DM}(\phi
= 0.15) \sim 27.5^{\circ} C$. In agreement with visual
observations (section \ref{Volumetric determination of the phase
diagram}), the isotropic phase of $5DH \mu em$ extends to
temperatures lower than the bulk $T_{NI}$. By fitting $B_d$ to
$A_{5CB}(1-\phi)/(T-T^*)$, we obtain a value for $T^*$ depending
on $\phi$, namely $T^*(\phi=0.075)= 27.3^{\circ} C$ and
$T^*(\phi=0.15)= 21.8^{\circ} C$. As $\phi$ increases, the maximum
value attained by the Kerr constant decreases, in agreement with
the increased difference between $T_{DM}$ and $T^*$. These data,
which will be discussed in Section \ref{me LS} together with the
light scattering results, indicate that the electro-optic
susceptibility of the mixture is dominated by the LC component and
that its behavior is simply the one expected for an isotropic LC
approaching the nematic transition with shifted $T^*$.

\begin{figure}
\begin{center}
\includegraphics[width=0.8\textwidth]{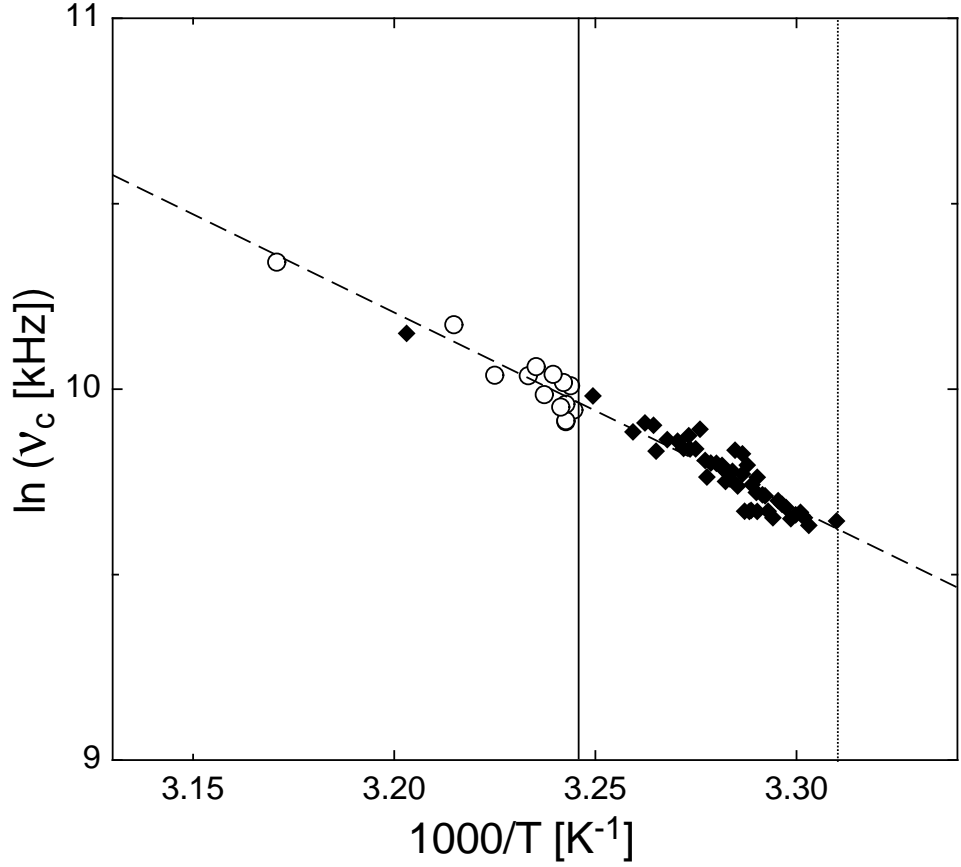}
\caption{Arrhenius plot of $\nu_c$ vs. $1000/T$ for bulk 5CB
(circles) and for $5DH \mu em$ sample with $\phi=0.15$ (diamonds).
Dashed line: Arrhenius fit of all data. Vertical lines indicate
$T_{NI}$ (solid line) and $T_{DM}(\phi=0.15)$ (dotted line).
\label{figura9}}
\end{center}
\end{figure}

Since the individual molecular dynamics is typically unaffected by
the onset of paranematic fluctuations \cite{pre22}, as also
confirmed by our data for bulk 5CB, it is expected that the
characteristic frequencies extracted from EBS $5DH \mu em$ would
also not be influenced by pre-demixing phenomena. In Fig.
\ref{figura9} we plot  $\nu_c$ vs. $1/T$ for the comparison with
bulk results. Within the experimental uncertainty shown by the
spread in the data point, the Arrhenius behavior of the $5DH \mu
em$ $\nu_c$ data coincides with the bulk one, indicating that the
activation energy for molecular reorientation is equal in the two
systems. Thus the presence of the micelles doesn't affect the
molecular environment experienced by the LC molecules, which
remains basically isotropic-like down to $T_{DM}$. Given the large
fraction of 5CB molecules in contact with the DDAB molecules
around the micelles (an estimate will be given in Section
\ref{discussion}), this result clearly indicates that no
significant anchoring of LC molecules is induced at the micelle-LC
interface.

\begin{figure}
\begin{center}
\includegraphics[width=0.8\textwidth]{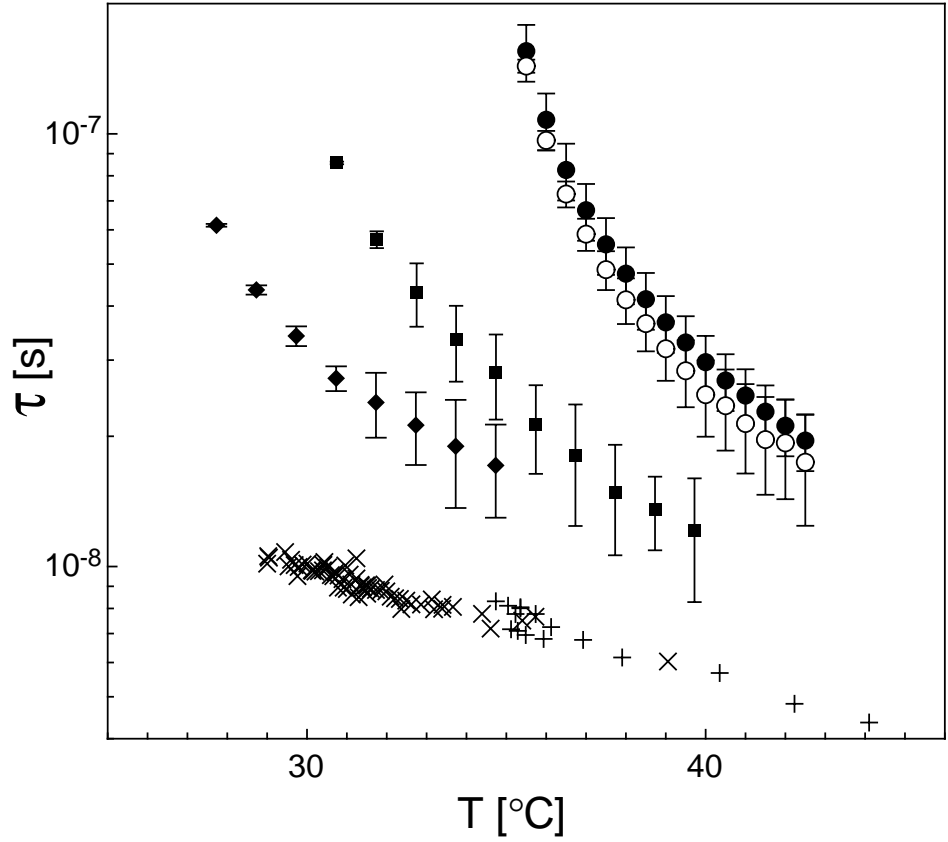}
\caption{Comparison between characteristic times and inverse
frequencies characterizing the dynamic light scattering and
electric birefringence spectra. Full dots: $\tau_{PN,VV}$ for bulk
5CB. Empty dots: $\tau_{PN,VH}$ for bulk 5CB. Squares:
$\tau_{PN,VH}$ for the $\phi=0.075$ $5DH \mu em$. Diamonds:
$\tau_{PN,VH}$ for the $\phi=0.15$ $5DH \mu em$. Pluses: $1/(2 \pi
\nu_c)$ for bulk 5CB. Crosses: $1/(2 \pi \nu_c)$ for the
$\phi=0.15$ $5DH \mu em$ \label{figura10}}
\end{center}
\end{figure}

In Fig. \ref{figura10} we show the characteristic time $\tau_c= (2
\pi \nu_c)^{-1}$ extracted from the data in Fig. \ref{figura9}, to
be compared with the characteristic time for paranematic
fluctuations, as discussed below.

\subsection{Light Scattering Results: Extracting Paranematic and
Micellar Contributions}\label{Light Scattering Results: Extracting
Paranematic and}

\subsubsection{Bulk 5CB}\label{Bulk 5CB LS}

Light scattering provides valuable information on both the statics
and dynamics of pretransitional paranematic fluctuations through,
respectively, the intensity and the correlation function of the
scattered light.
Because of the first order nature of the I-N phase transition,
light scattering results are usually accurately described by the
mean field theories. Accordingly, the scattered intensity, as
described by the Rayleigh ratio R, grows as \cite{pre22}

\begin{equation}
R=\frac{k_B T}{a(T-T^*)} \label{intvst}
\end{equation}

while the correlation time of the paranematic fluctuations is
given by \cite{pre22}

\begin{equation}
\tau_{PN} = \frac{k_B T \eta }{a(T-T^*)} \label{tausuetavst}
\end{equation}

where $\eta$  is the viscosity.

Our bulk data are in agreement with the mean field description and
with previous investigations. Measured $I_{VV}$ and $I_{VH}$
intensities are well fit by Eq.\ref{intvst} with added a
background constant, accounting for stray light. Fig.
\ref{figura6} shows such a fit for $I_{VH}$, yielding $T^* =
33.8^{\circ} C$ (and thus $T_{NI} - T^* = 0.9^{\circ} C$), in
acceptable agreement with literature \cite{pre17} and EBS data
analysis. Polarized and depolarized intensity are found to be
proportional, as predicted \cite{pre22}:

\begin{equation}
\langle I_{VH} \rangle= \frac{3}{4} \langle I_{VV} \rangle
\label{ivvivh}
\end{equation}

This is shown in Fig. \ref{figura5} where cross symbols represent
$\frac{4}{3} \cdot \langle I_{VH} \rangle$ and nicely overlay on
$\langle I_{VV} \rangle$ data.

The correlation functions obtained in the bulk can be well fit
with exponential decays. The analysis of the $VV$ correlation
functions yields to values of $\tau_{PN}$ (see Fig. \ref{figura10},
full dots) about 15
\% larger than those obtained from VH scattering. This observation
is in line with prediction in the frame of mean field analysis as
a consequence of coupling between shear viscosity and density
fluctuations, which bears detectable effects in the VH Rayleigh
spectra \cite{pre22}. On the basis of equation (23) of Ref.
\cite{pre22} and by using values obtained from experiments on
MBBA, we would predict $\tau_{PN,VV}/ \tau_{PN,VH} \sim 1.3$,
about double of what observed in 5CB.

$\tau_{PN}/ \eta$  can be well approximated by $(T-T^*)^{-1}$, as
predicted by Eq. \ref{tausuetavst} and shown in Fig.
\ref{figura11}. The extracted divergence paranematic temperature
is $T^* = 34.4^{\circ} C$ is too high with respect to the one
obtained from $I_{PN}$ in the same light scattering experiment.
The origin of this difference appears to be seeded
in the behavior of $\eta$ \cite{pre16}.

\subsubsection{$5DH \mu em$}\label{me LS}

Microemulsion data require some additional analysis. The measured
correlation functions (Fig.\ref{figura7}) very clearly demonstrate
the presence of two distinct processes: a fast one, appearing in
both VV and VH correlations, clearly due to paranematic
fluctuations, and a slower one, appearing only in the VV
correlations, detectable only in the microemulsions. This last
component, indicating the presence of aggregates within the 5CB
matrix, is easily interpreted as intensity scattered by the
inverted micelles. Thus the total scattered intensity $I_{TOT} = I_{PN} +  I_{M}$
is the incoherent sum of intensity scattered by LC orientational
fluctuation ($I_{PN}$) and by micelles ($I_{M}$). Specifically

\begin{equation}
I_{VV} \equiv I_{TOT,VV} = I_{PN,VV} + I_{M}
\end{equation}

while

\begin{equation}
I_{VH} \equiv I_{TOT,VH} = I_{PN,VH}
\end{equation}

The large difference between the correlation times of the two
components makes it possible to identify them. Let's indicate with
$\tau_{mid} \sim 1 - 10 \mu s$ a correlation time long with
respect to paranematic fluctuation time and short with respect to
micellar diffusional times. Because of the statistical
independence of the two terms we have

\begin{equation}
\begin{array}{ll}
G_{2,VV}(0)-G_{2,VV}(\tau_{mid}) &= c ( \langle I_{PN,VV}^2
\rangle + 2 \langle I_{PN,VV} \rangle
\langle I_M \rangle )\\
G_{2,VV}(\tau_{mid})-G_{2,VV}(\infty) &= c \langle I_M^2 \rangle
\end{array}
\end{equation}

where $c$ is a factor related to the collection optics. Combining
the two equations,  $\langle I_M \rangle$ can be derived from
the total polarized scattered intensity and from the decay of the correlation
functions. Alternatively,  $\langle I_M \rangle$
can be extracted from the polarized scattered intensity $\langle
I_{VV} \rangle$ as

\begin{equation}
\langle I_M \rangle= \langle I_{VV} \rangle -\frac{4}{3} \langle
I_{VH} \rangle
\end{equation}

where we have used the notion that  $\langle I_{PN,VV}\rangle= 4/3
\langle I_{VH} \rangle$ as observed for paranematic scattering in
bulk 5CB (eq. \ref{ivvivh}). Fig. \ref{figura12} compares the
values of $\langle I_M \rangle$ vs. T obtained from the two
procedures. The comparison clearly confirms that the T dependence
of  $\langle I_M \rangle$  is indeed reliably extracted. The
determination of $\langle I_M \rangle$ from DLS data, however,
requires extracting $G_{2,VV}(0)$ from the correlation functions.
This has been done by fitting $G_{2,VV}(\tau)$ by a sum of two
exponentials (see Fig. \ref{figura7}) enabling determining
$G_{2,VV}(0)$, as well as $\tau_{PN}$ and the micellar correlation
time $\tau_M$. $\tau_{PN}$ obtained from $G_{2,VH}(\tau)$ are more
reliable than those obtained from $G_{2,VV}(\tau)$ since in this
last case the intensity scattered by the micelles, basically
constant on the times scale of paranematic fluctuations, beats
with the light diffused by orientational fluctuations thus
introducing an heterodyne contribution to the correlation
function. Therefore, $\tau_{PN}$ extracted from $G_{2,VV}(\tau)$
would, in principle, require a compensating correction,
unnecessary in the VH scattering. Although in practice such a
correction is small since $I_M / I_{LC}  <  15 \%$ always, we will
in the following refer to $\tau_{PN}$ obtained from
$G_{2,VH}(\tau)$.

In summary, light scattering measurements enable extracting
$\langle I_{PN} \rangle$ , $\langle I_M \rangle$ ,  $\tau_{PN}$,
$\tau_M$ as a function of T and for the different samples. In what
follows we will separately discuss results pertaining the two
sides of the $5DH \mu em$ behavior: LC in the presence of
nanoinclusions and micelles within a solvent with ordering
fluctuations.

Inspection of  $\langle I_{PN} \rangle$ vs. T measured in the $5DH
\mu em$ shown in Fig. \ref{figura6} reveals that: (i) at any given
T, paranematic fluctuations are depressed by the presence of the
micelles; (ii) an isotropic phase like behavior extends down to
$T_{DM}$, i.e. even at T lower than the bulk $T_{NI}$; (iii) the
maximum value of $\langle I_{PN} \rangle$ for each given
preparation, obtained for $T = T_{DM}$, is a decreasing function
of $\phi$.

By fitting $I_{PN}(T)$ as for the bulk data we obtain $T^*( \phi =
0.075) = 27.2^{\circ} C$ and $T^*( \phi = 0.15) = 21.9^{\circ} C$,
closely matching the results obtained from EBS. These results are
reported in the phase diagram in Fig. \ref{figura2}, where the
dotted line interpolating the data guesses the behavior of
$T^*(\phi)$.

\begin{figure}
\begin{center}
\includegraphics[width=0.8\textwidth]{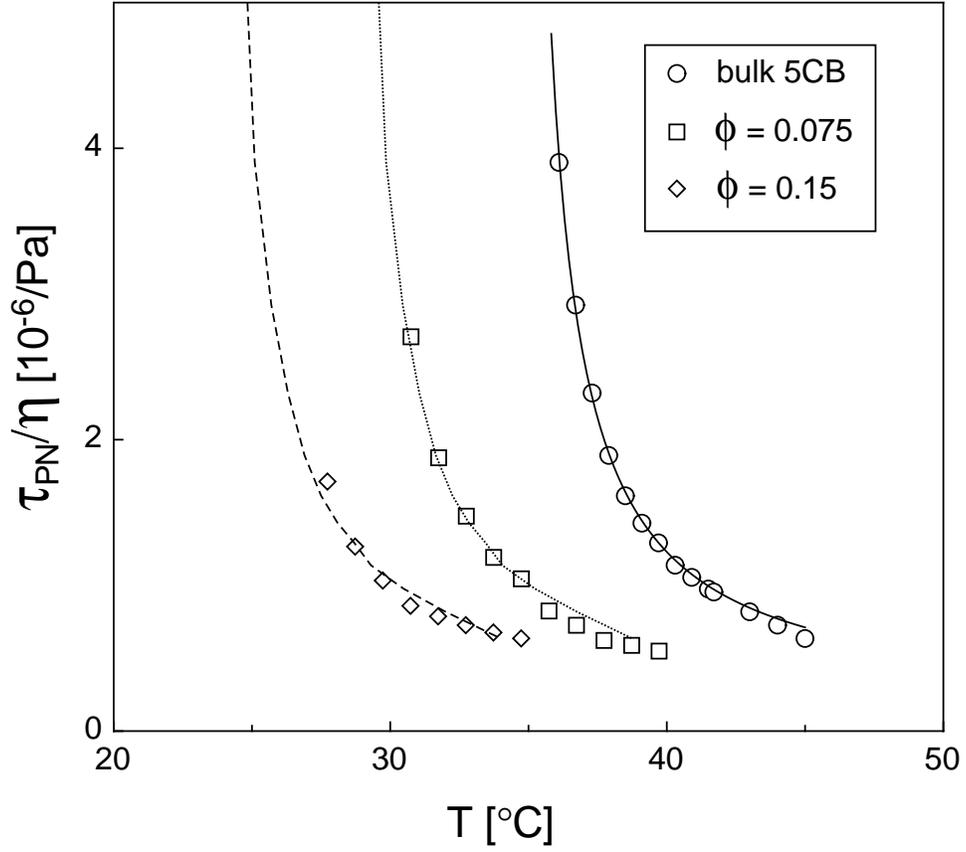}
\caption{\label{figura11} Paranematic correlation time $\tau_{PN}$
divided by the shear viscosity $\eta$ for bulk 5CB (circles) and
$5DH \mu em$ with $\phi=0.075$ (squares) and $\phi=0.15$
(diamonds) as a function of temperature.}
\end{center}
\end{figure}

$\tau_{PN}/ \eta$  data vs. T are shown in Fig. \ref{figura11}.
Given the uncertainty on the T dependence of the viscosity, as
discussed above, we prefer not to fit the data, but to evaluate
the magnitude of the temperature shift induced by the presence of
the micelles by finding the best overlap between the $5DH \mu em$
data and the bulk 5CB data shifted down in temperature of $\delta
T$ as shown in Fig. \ref{figura11}. Best overlaps have been
obtained with  $\delta T = 6.25^{\circ} C ( \phi = 0.075)$ and
$\delta T = 11.0^{\circ} C ( \phi = 0.15)$, to be compared,
respectively, with $T^*(bulk)-T^*(\phi = 0.075) = 6.6^{\circ} C$
and with $T^*(bulk)-T^* (\phi = 0.075) = 11.9^{\circ} C$ as
obtained in static LS.

The overall agreement between the $T^*$ from static and dynamic LS
and from the low frequency Kerr measurements, supports the notion
that the annealed disordering provided by the micelles is
affecting the LC behavior simply by shifting it to lower
temperatures and by increasingly precluding the development of
paranematic fluctuations by the appearance of the demixing
transition, i.e. $T_{DM}( \phi = 0.15) - T^*( \phi = 0.15) [\sim
10^{\circ} C] > T_{DM}( \phi = 0.075) - T^* ( \phi = 0.075) [\sim
6^{\circ} C] > T_{NI} - T^* (bulk) [\sim 1^{\circ} C]$.

\subsubsection{Micellar behavior}\label{Micellar behavior}

\begin{figure}
\begin{center}
\includegraphics[width=0.8\textwidth]{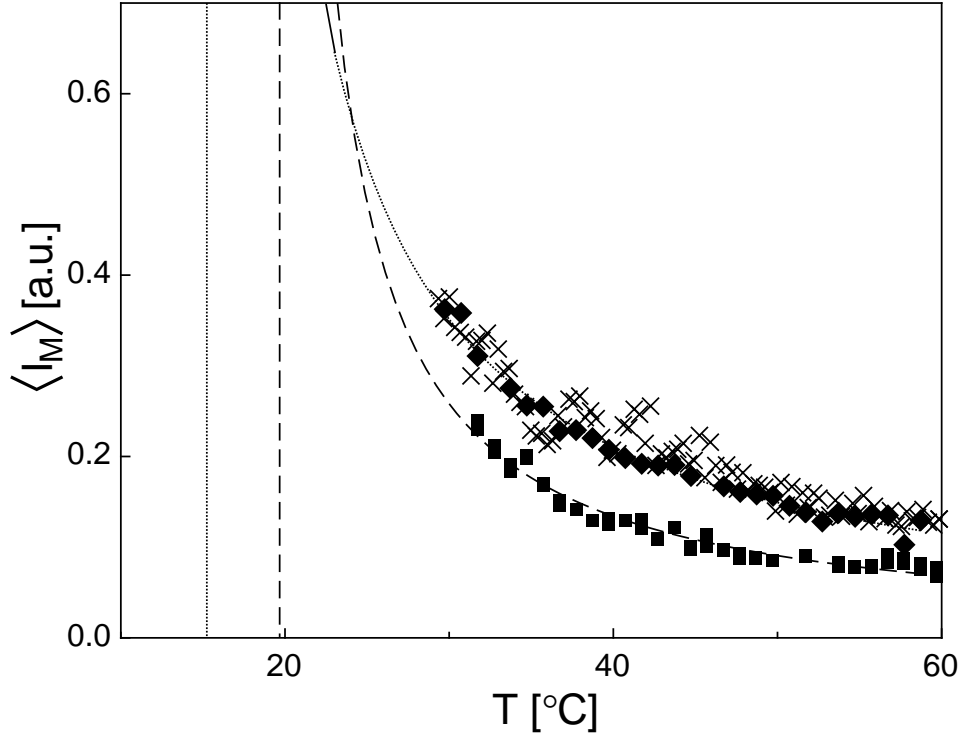}
\caption{\label{figura12}Intensity scattered by microemulsion
droplets $\langle I_{M} \rangle$ extracted from dynamic light
scattering data as a function of temperature. Squares:
$\phi=0.075$. Diamonds: $\phi=0.15$. Crosses: $\langle I_{M}
\rangle$ obtained from the difference between $I_{VV}(\phi=0.15)$
and $4/3 \cdot I_{VH}(\phi=0.15)$ and normalized to match diamonds
at $T=60^{\circ}C$. Lines: fittings with $(T-T_S)^{-1}$ yielding
$T_S(\phi=0.075)=19.4^{\circ}C$ (dashed vertical line),
$T_S(\phi=0.15)=14.6^{\circ}C$ (dotted vertical line)}
\end{center}
\end{figure}

\begin{figure}
\begin{center}
\includegraphics[width=0.8\textwidth]{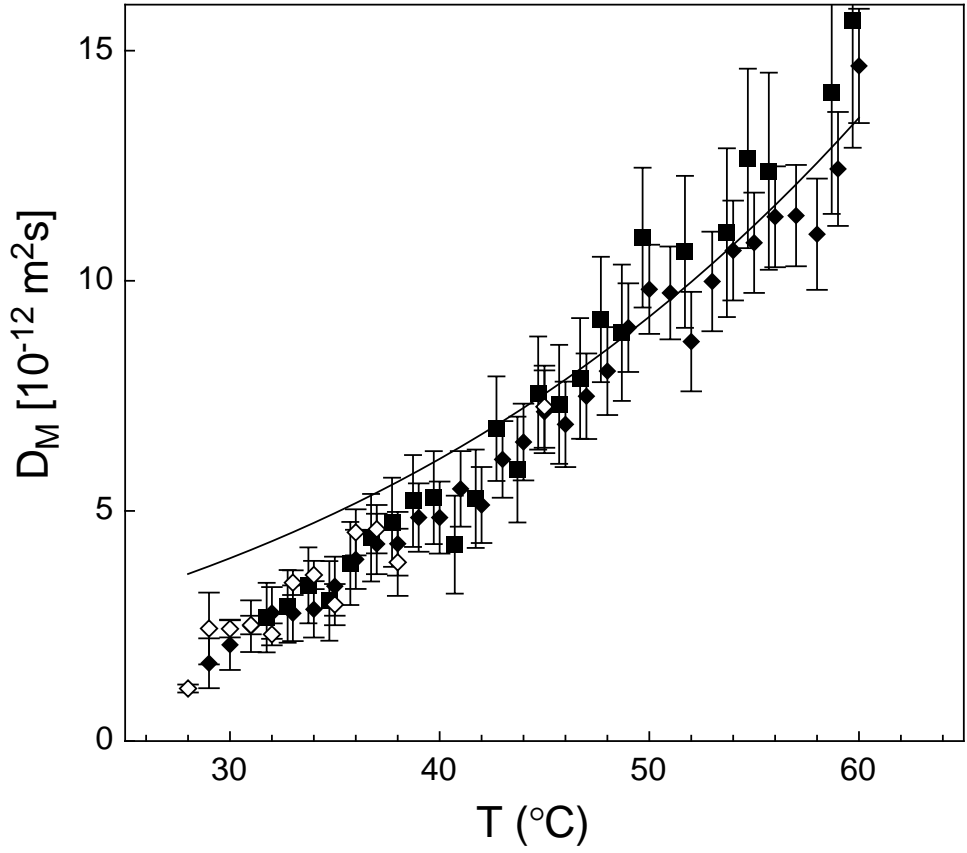}
\caption{\label{figura13}Micellar diffusion coefficients $D_M$ as
a function of temperature for $5DH\mu em$ samples with $\phi =
0.15$ (full diamond), $\phi = 0.075$ (squares). Empty diamonds
$D_M$ measured in the upper phase of a $\phi = 0.075$ $5DH\mu em$
at $T=27^{\circ}C$ where independent estimates indicate a micellar
concentration of $\phi = 0.14 $. Line: expected temperature
dependence of $D_M$ for non interacting microemulsion droplets
having radius $R=18 \Ang$ calculated on the bases of $\eta(T)$
from Ref. \cite{pre20}.}
\end{center}
\end{figure}

The temperature dependence of  $\langle I_M \rangle$  for the two
concentrations is shown in Fig. \ref{figura12}. The most
interesting characteristic of these data is the growth of $\langle
I_M \rangle$  as T decreases, which cannot be understood simply in
terms of T dependence of refractive indices, which, in the T
interval of Fig. \ref{figura12}, vary at most by 1 \%. Analogous
deviation from simple behavior is observed in the diffusion
dynamics of the micelles. The micellar diffusion coefficient $D_M$
extracted from  $\tau_M^{-1} = 2D_M q^2$ (where $q$ is the
scattering vector) is shown in Fig. \ref{figura13} and compared
with the $D_M$ which would be expected from the Einstein equation
$D_M = (k_B T)/(6 \pi \eta R)$ in the case of spheres of radius $R
= 18 \Ang$ diffusing in the isotropic phase, where the viscosity
is taken from Ref. \cite{pre20}. The radius has been chosen so to
approximate the observed $D_M$ at the highest measured
temperatures. As evident from the comparison, in the proximity of
the demixing transition, the measured dynamics is slower than the
expected one by a factor of about 2 in both the $ \phi = 0.15$ and
the $\phi = 0.075$ samples.

The comparison between $D_M(T,\phi )$ and $(k_B T)/(6 \pi R \eta
(T))$ is based on the assumption that the local viscosity
experienced by inverted micelles in the $5DH \mu em$ is unchanged
with respect to bulk 5CB and that, for $T < T_{NI}$, where no bulk
data for isotropic viscosity exists, $\eta(T)$ is given by the
same Arrhenius dependence as in the bulk. This assumption is is
based on our EBS results on the Kerr $\nu_c$, which proved that
the local molecular environment in the isotropic phase of $5DH \mu
em$ overlaps the low T extrapolation of the bulk Arrhenius. Given
that viscosity and molecular diffusion share their basic physical
origin \cite{pre23}, we conclude that similar T dependence is
expected for $\eta (T)$. Use of another among the published 5CB
viscosities would not change the comparison in Fig.
\ref{figura13}. Namely, by using $\eta (T)$ as in Ref.
\cite{pre19} we would obtain $R = 21 \Ang$ and a maximum ratio
between expected and measured D of about $2.1$, while by using
$\eta (T)$ as in Ref. \cite{pre21} we would obtain $R = 24 \Ang$
and a maximum ratio between expected and measured $D_M$ of about
$1.8$. Given this dispersion of the R values, in the following
discussion we will assume $R=20 \pm 2 \Ang$.

According to our data, and within our experimental error, the
diffusion coefficients measured for the two $5DH \mu em$ micellar
concentrations vs. T are basically the same (see Fig.
\ref{figura13}). Although this result may appear as a
straightforward indication of having droplets of the same size in
the two samples, some analysis is instead required. In fact, $\phi
= 0.15$ is a rather significant volume fraction, at which effect
from interparticle interactions may be expected. When expressed
through the virial expansion \cite{pre35}, the collective
diffusion coefficient measured at a given particle concentration
is

\begin{equation}
D_M=D_{M,0}(1+k_D \phi) \label{viriald}
\end{equation}

where $D_{M,0}$ is the diffusion coefficient at infinite dilution
and $k_D$ is the virial coefficient expressing the correction to
collective diffusion arising from interparticle interactions.
According to the data here presented, the product $k_D \phi$ assumes
the same value in the two different samples, thus indicating a
$\phi$ independent D, and possibly $k_D = 0$.

In order to check the robustness of this evidence and also to
explore possible deviations from simple one-component Maxwell rule
for the phase coexistence, we have studied the upper phase -
denser in micelles - obtained by letting a  $ \phi = 0.075$ sample
separate at $T = 27.5^{\circ} C$. The resulting micellar volume
fraction, calculated from the volume ratio of the phases and
assuming a 2 \% micelles in the lower N phase (see Fig.
\ref{figura2}) was $\phi = 0.14$. Dynamic light scattering
measurements on the upper isotropic phase are possible because the
time for the two coexisting phases to remix is much larger than
the time to perform the experiment. The micellar diffusion
coefficient obtained in this way is compared in Fig \ref{figura13}
with the $D_M$ measured in sigle phase $5DH \mu em$. The overlap
of the data confirms the absence of significant concentration
dependence of $D_M$ at least within our experimental error.
Moreover it supports the single component approximation,
previously assumed in drawing the phase diagram (see Section
\ref{Volumetric determination of the phase diagram}).

The deviation of $D_M(T)$ from $k_B T/ \eta(T)$ indicates that the
product $k_D \phi$ , constant for $T > 40^{\circ}C $, grows
appreciably as the system approaches the demixing transition.
Altogether, the matching of $D_M(T)$ with $k_B T / \eta(T)$ at
high T, and the independence of $k_D \phi$ on $\phi$ , strongly
suggests that, at high enough temperature, $k_D \sim 0$.
Vanishing virial coefficients are not at all a surprise for water in oil
microemulsions. Systematic investigations in systems prepared with
different choices of oil and surfactants have revealed that
inverted micelles may interact with a wide range of variable
strengths, ranging from negative - hard sphere like - to positive
virial coefficient \cite{pre25}. This is reflected in other
studies too. In one of them \cite{pre26}, it was found that a
large body of data could be satisfactorily understood on the basis
of hard sphere repulsion, with the hard sphere radius few $\Ang$
smaller than the measured hydrodynamic radius. Other
investigations \cite{pre27,pre28} concluded that contact
attractive interactions had to be included to interpret the data.
In another study \cite{pre29} it was actually found that
collective diffusion was independent on droplet concentration, as
in the $5DH \mu em$ case here. The existence of attractive
interactions in inverted micelles is usually attributed to an
entropic gain coming into play when the hydrophobic tail of the
surfactants of two droplets overlap \cite{pre30}. Our data
indicate that such a "stickiness" interaction may be present in
our system to compensate, as far as $k_D$ is concerned, for the
hard sphere repulsion of the hydrophilic cores, certainly present.
Thermal dependence of such interpenetration interaction also
explains the inverted consolution curve ("lower" critical solution
temperature) observed in the water-AOT-octane \cite{pre30,pre31}.
We stress that such inverted demixing (induced upon increasing T)
is the only gas-liquid like phase separation observed in
water-in-oil microemulsions and that its reversed nature clearly
demonstrates that the demixing observed in $5DH \mu em$ (induced
upon lowering T) cannot be explained by the same attractive
potential.

Previous works on DDAB \cite{pre32}, focused on w/o
microemulsions prepared by using hydrocarbon oils and
with a much higher water/DDAB concentration ratio, propose a
calculation scheme for the micellar size which would yield, in our
case $R \sim 15.5-19.5 \Ang$, depending on how extended the DDAB
chains are. Care should be taken, though, since
the radius may depend on the nature of the oil. By comparing phase diagrams obtained with
different kind of oils, two interesting elements appear: (i) DDAB
is insoluble in alkanes if no water is added, indicating the
existence of a minimum curvature of the DDAB layer, which also
appears to be dependent on the alkane chain length
\cite{pre32,pre33}; (ii) DDAB is soluble in toluene with no added
water but it does not form micelles \cite{pre24}. DDAB is soluble
in the isotropic phase of 5CB with no added water. However, DLS
measurements have revealed that, at variance with the toluene
system, DDAB in the isotropic phase of 5CB form micelles. Their
radius, measured in a sample with 15 \% DDAB in 5CB by a different
DLS set-up equipped by a more powerful laser source, is $R \sim
18 \Ang$. This number should be considered as very approximate
because no analysis has been done on possible intermicellar
interactions. Overall, the solubility properties of DDAB in 5CB
appears to be quite different from that in both alkanes and
toluene.

A simple estimate of the $5DH \mu em$ inverted micellar structure
is as follows. By assuming an intermediate (not fully extended)
DDAB hydrocarbon chain length of $14 \Ang$ \cite{pre32}, it
follows that the water core radius $R_W \sim 6 \Ang$. Since the
DDAB/water volume ratio extends to the micellar structure, should
the DDAB not compenetrate with 5CB molecules but rather form a
compact spherical shell, we would have micelles of $13.5 \Ang$.
This fact suggests a very strong interfingering of DDAB and 5CB
molecules, in agreement with what implied by the solubility of
DDAB in 5CB and by the large curvature of DDAB in 5CB. Overall,
the DDAB volume in every droplet is small, and corresponds to
about 13-14 DDAB molecules.

\subsubsection{Micellar concentration fluctuations}\label{Micellar concentration fluctuations}

The growth of  $\langle I_M \rangle$  and of $D(T) \eta(T) / k_B
T$ as $T_{DM}$ is approached could, in principle, be explained in
a variety of manners. It could be due to the growth of a 5CB layer
of radially anchored molecules on the micelles surface, thickening
as a consequence of the increasing nematic correlation length.
Alternatively, it could follow from restructuring of the
spontaneously assembled water-DDAB droplets. Finally, it could
reflect the onset of attractive intermicellar forces. We will
argument in favor of this last explanation.

The development of a radial nematic layer around the micelles
could certainly bring about a slowing down of their diffusion, but
it certainly would not increase their scattering cross section for
light. Indeed, given the size of the particles - well within the
Rayleigh scattering regime \cite{pre34}, isotropically anchored
molecules cannot be distinguished from a layer of isotropically
distributed unanchored molecules.

Restructuring of the self-aggregated nanodroplet could take place
but only allowing for non-spherical shapes, since the ratio
[DDAB]/[water] has to be maintained in any aggregate. In order to
account for a growth of about a factor 3 in the scattered
intensity, aggregates should increment their volume of about 3
times. This would lead to the formation of rod-like micelles whose
diffusion coefficient can be explicitly calculated and results to
be at least 4 times larger than expected for spheres of constant
radius. Since the observed slowing down is only of a factor 2, we
don't see room for pursuing this explanation.

Interparticle attraction, besides slowing down their diffusion,
also give rise to concentration fluctuations, in turn enhancing
the amount of scattered light.  This effect is described by the
second virial coefficient $k_I$ in the virial expansion

\begin{equation}
I_M= I_{M,0}(1+k_I \phi)^{-1}
\end{equation}

where $I_{M,0}$ is the scattered intensity measured if
interactions were not present. $k_D=k_I-k_f$, where both
$k_D$ and $k_f$ are suitable integrals of the pair
potential \cite{pre35}. While $k_I$ describes concentration
fluctuations, $k_f$ takes into account hydrodynamic interactions.
Attractive interactions correspond to negative coefficients. For
short ranged potentials, $k_f$ has typically the same sign of
$k_I$ but is smaller in amplitude (about half or less). $\langle
I_M \rangle$ data in Fig. \ref{figura12} are extracted from
$\langle I_{VV} \rangle$ data on the basis of the ratio of slow
and fast dynamics in the intensity correlation functions. Thus, as
in the case of  $I_{VV}$ , the values of  $\langle I_M \rangle$
cannot be compared between different samples. However, the fact
that  $k_D \sim 0$ at high
temperature, suggests that also $k_I$ is negligible in the same
temperature range, as found in other systems having vanishing
$k_D$ \cite{pre35}. The growth of $\langle I_M \rangle$ as T
decreases indicates that $k_I$ is temperature dependent, ranging
from its "high T" value $k_I^{HT}=0$, to a value at the demixing
temperature $k_I \sim -4.3$ in the case of $\phi=0.15$ and a value
$k_I \sim -9.3$ in the case of $\phi=0.075$ (from the data in Fig.
\ref{figura12}).

Since an increment in the attractive potential implies an
increment in the negative value of both $k_I$ and $k_f$,
with the second smaller
in modulus than the first, we expect $k_D$ to be negative and $k_D
> k_I$. This prediction is in agreement with the experimental
result, in which the effects of interactions are more conspicuous
in the scattered intensity than in the diffusion coefficient.
Therefore, while neither the growth of a nematic layer around the
droplets nor micellar restructuring results compatible with
experimental results, the appearance of intermicellar attraction
is capable to account for the data.

Although the specific nature of intermicellar interactions is not
experimentally accessible, their occurrence in the proximity of
the demixing transition, in a temperature interval in which
orientational fluctuations also develop, suggests the existence of
a connection between the two phenomena. It is exactly to explore
such a connection that we have developed the model described in
the following Section.

\section{Theoretical model}\label{Theoretical model}
The nematic phase can be modelled in a variety of ways, having
various degrees of realism. Striking, for its simplicity and
effectiveness in capturing the strength of the first order I-N
phase transition, is the Lebwhol-Lasher (LL) model \cite{pre40}, a
spin lattice model with the simplest orientational coupling
compatible with the non-polar nematic symmetry. This model is also
interesting because it enables studying the effects of quenched
disorder once a random field is properly added \cite{pre47,pre48}.
Hence, the LL model appears as an attractive
starting base to investigate the effect of nanoimpurities.
The LL model can also be reckoned as the lattice counterpart of
the widely known Maier-Saupe model \cite{pre57,pre58}, which has
been repeately studied ever since \cite{pre59}.

The analysis of the pretransitional behavior of the pure LL
indicates that, if a mapping is attempted between spins and
molecules, a single spin represents not a single molecule but
rather a cluster of about 50 molecules \cite{pre49}. In the case
of $5DH \mu em$, the droplets in the solution are comparable in
sizes with the length of the LC molecules, while filling the
volume of 20-40 5CB molecules. Hence, one is then led to suspect
that a lattice model containing a mixture of LC and micelles,
taken to be of the same size, might be sufficient to capture the
main qualitative features of the experiments presented so far. We
now show that this is indeed the case. In our model, LC are
represented by pointless unit vectors $\hat{s}$ which are free to
rotate in three dimensional space, but are fixed in position to a
lattice of spacing $a$:

\begin{equation}
\hat{s} = (s^1 , s^2 , s^3)=(\sin{\theta} \cos{\phi} ,
\sin{\theta} \sin{\phi} , \cos{\theta} ) \label{theor1}
\end{equation}

Micelles, on the other hand, are simply mimicked by vacancies, and
are then represented - in the spirit of a lattice gas model - by a
lattice variable $\sigma =0(,1)$ when the lattice site
is vacant (occupied by LC).\\

As usual, it proves convenient to introduce the traceless
quadrupole tensor

\begin{equation}
    Q^{\alpha\beta}(\hat{s})=\frac{1}{2}(3s^\alpha s^\beta -
    \delta^{\alpha\beta} ) \qquad \qquad \alpha,\beta=1,2,3
\label{theor2}
\end{equation}

The final Hamiltonian then reads:

\begin{equation}
  -\beta H = K_{qq}
\sum_{\alpha \beta} \sum_{\langle xy \rangle} Q_{x}^{\alpha\beta}
Q_{y}^{\beta\alpha}
  \sigma_x \sigma_y +
  K_{\sigma\sigma} \sum_{\alpha \beta} \sum_{\langle xy \rangle} \sigma_x \sigma_y\\ +
  \sum_{\alpha \beta} \sum_{x}h_x^{\alpha \beta} Q_x^{\alpha \beta} \sigma_x - \sum_x
  \beta \mu \sigma_x
\label{theor3}
\end{equation}

Here the first term reflects the fact that two LC have the general
tendency to align along a common direction with coupling $K_{qq} =
\beta J_{qq}$, $(\beta = 1/k_B T)$, the second term is a standard
lattice gas Ising-like interaction with coupling $K_{\sigma
\sigma}=\beta J_{\sigma \sigma}$, and the next two terms
correspond to the external fields ($h^{\alpha \beta}$ for the LC
and $\mu$ for the micelles) controlling the  LC orientation and
micelle concentrations respectively.

In the following the average fraction of micelles will be denoted
as $\phi$. It is worthwhile noticing that in Eq
(\ref{theor3}) the micelles interact both directly (through a
$K_{\sigma \sigma}$ coupling) {\it and} indirectly, because of the
presence of the LC. As we shall show, the latter are responsible
for the most interesting features of the mixture, and are present
even in the absence of the former.

\subsection{Mean Field Analysis}\label{Mean Field Analysis}
The diluted LL model has been studied using Monte Carlo
simulations \cite{pre60,pre61} In the following we shall study it
within a mean field approximation akin that of its continuum
counterpart \cite{pre59}.

To this aim we define a trial Hamiltonian

\begin{equation}
- \beta H_0 = \sum_{\alpha \beta} \sum_{x} b_x^{\alpha \beta}
Q_x^{\alpha \beta} \sigma_x + \sum_x \lambda_x \sigma_x
\label{theor4}
\end{equation}

where $b^{\alpha \beta}$ and $\lambda$ are variational parameters
for the LC and micelles respectively. As remarked, at variance of
Ref. \cite{pre7}, we shall exploit an Hamiltonian which is not
simply factorized in two terms which depend on one variable only.
Clearly a more general choice
\begin{equation}
\tilde{H}_0 = \sum_x h \left(\{ Q \sigma_x\} \right)
\end{equation}
($h\left(\{ Q \sigma_x \} \right) $ being a single site
hamiltonian) could be exploited, but the calculations quickly
become analytically unmanageable.

Accordingly we shall use the following order parameters

\begin{eqnarray}
q_x^{\alpha \beta}&=& \left \langle Q_x^{\alpha \beta} \sigma_x
\right \rangle \\ \nonumber n_x &=& \left \langle \sigma_x \right
\rangle \label{theor5}
\end{eqnarray}

where averages are over the trial Hamiltonian. Note that
$\phi=1-n$.

The calculation now follows standard procedures. Once the
mean field free energy $\beta F_{MF}$ has been obtained, it proves
convenient to Legendre transform to a mixed function

\begin{equation}
\beta \Gamma_{MF} = \beta F_{MF} + \sum_x h_x^{\alpha \beta}
q_x^{\alpha \beta} + \sum_x \mu^*_x n_x \label{theor6}
\end{equation}

where $\mu^* =\beta \mu$. This function clearly depends upon the
variational parameters $b^{\alpha \beta}$ and $\lambda$, whose
explicit values are obtained by the requirement that the free
energy is stationary at those values. The final results for this
mixed free energy  in terms of the homogeneous value of the
micelles order parameter $n$ and of the LC order parameter $q$ as
identified by the diagonal matrix

\begin{equation}
q^{\alpha \beta} = \begin{vmatrix} -\frac{q}{2} & 0 & 0 \\ 0 &
-\frac{q}{2} & 0
\\ 0 & 0 & q \end{vmatrix}
\label{theor7}
\end{equation}

is

\begin{eqnarray}
\frac{\beta \hat{\Gamma}_{MF}(q,n)}{N} &=& \frac{9}{2} K_{qq} q^2
- 3 K_{\sigma \sigma} n^2 + \frac{9}{2} K_{qq} q n +\\ \nonumber
&& + n \log{n} + (1-n) \log{(1-n)} - \log{4 \pi} -n
\log{\left[I_0\left(\frac{27}{2} K_{qq} q \right)\right]}
\label{theor8}
\end{eqnarray}

where we have defined the integrals

\begin{equation}
I_n(\lambda) = \int_0^1 dx ~ x^n e^{\lambda x^2} \label{theor9}
\end{equation}

\subsubsection{The phase diagram}\label{The phase diagram}

The isotropic-nematic transition is obtained by the conditions
(when $h^{\alpha \beta}=0$)

\begin{eqnarray}
\frac{\partial}{\partial q} \left[ \frac{\beta
\hat{\Gamma}_{MF}}{N} \right] &= 0\\ \nonumber
\frac{\partial}{\partial n} \left[ \frac{\beta
\hat{\Gamma}_{MF}}{N} \right] &= - \mu^* \label{theor10}
\end{eqnarray}

that is:

\begin{eqnarray}
q&=&n \Phi \left( \frac{27}{2} K_{qq} q \right) \label{theor11}
\end{eqnarray}
\begin{eqnarray}
\mu^*&=& 6 K_{\sigma \sigma} n - \frac{9}{2} K_{qq} q - \log
\left( \frac{n}{1-n} \right) + \log{\left[I_0 \left( \frac{27}{2}
K_{qq}q\right)\right]} \label{theor11}
\end{eqnarray}

where

\begin{equation}
\Phi ( \lambda) = -\frac{1}{2} + \frac{3}{2} \frac{I_2
(\lambda)}{I_0 (\lambda)} \label{theor13}
\end{equation}

Note that the Eq.(\ref{theor11}) is a self-consistency equation
for the nematic order parameter $q$.

It is instructive to compare the above results with the
corresponding one obtained in the absence of micelles, i.e. the
fully occupied Lebwohl-Lasher model (FLL), and with the simple
lattice gas case. The free energy for the former reads

\begin{eqnarray}
\left[\frac{\beta \hat{\Gamma}_{MF}(q)}{N}\right]_{FLL} &=&
\frac{9}{2} K_{qq} q^2 + \frac{9}{2} K_{qq} q - \log(4 \pi) -
\log\left[I_0\left(\frac{27}{2} K_{qq} q \right)\right]
\label{theor14}
\end{eqnarray}

where the order parameter satisfies
\begin{equation}
[q]_{FLL} = \Phi \left( \frac{27}{2} K_{qq} q \right)
\label{theor15}
\end{equation}

For the simple lattice gas (LG) case, on the other hand, we have

\begin{eqnarray}
\left[\frac{\beta \hat{\Gamma}_{MF}(n)}{N}\right]_{LG} &=& 3
K_{\sigma \sigma} n (1-n)+ n \log n + (1-n) \log(1-n)
\label{theor16}
\end{eqnarray}

\begin{equation}
[\mu^*]_{LG} = 6 K_{\sigma \sigma} n - 3 K_{\sigma \sigma}
-\log\frac{n}{1-n} \label{theor17}
\end{equation}

A couple of comments are here in order. First the $n \rightarrow
1$ limit of Eq. (\ref{theor7}) results into Eq. (\ref{theor14}) as
it should. Note that since $-1/2 \le Q^{\alpha \alpha} \le 1$, the
order parameter $q$ is bounded and hence $\mu^{*} \to \infty $
yields $n \to 1$. Second Eq. (\ref{theor8}) does not appear to be
a simple combination of Eq. (\ref{theor14}) and (\ref{theor16}) as
it has been argued on the basis of phenomenological theories
\cite{pre42}.

\begin{figure}
\begin{center}
\includegraphics[width=0.8\textwidth]{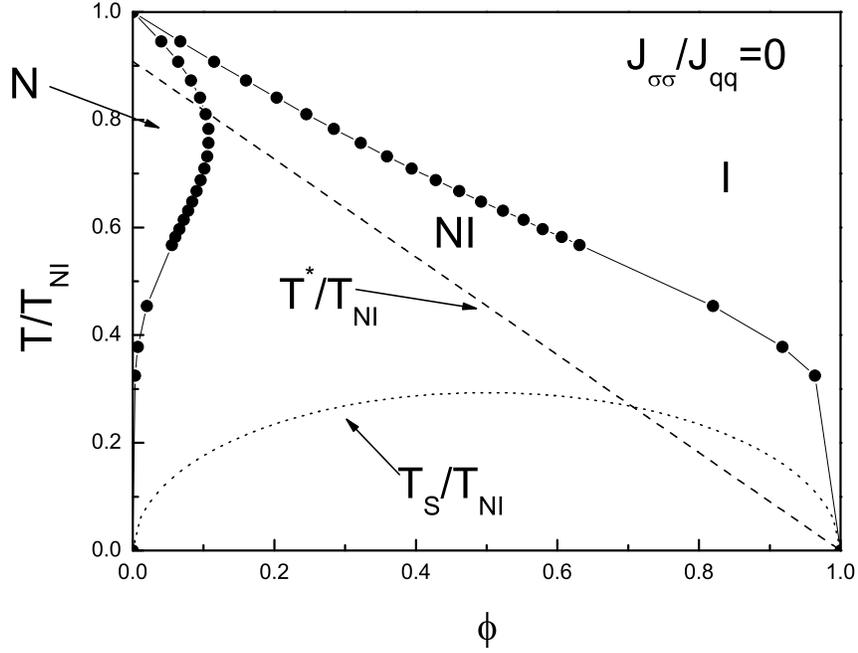}
\caption{Theoretical phase diagram calculated in the case of no
direct intermicellar interaction $J_{\sigma \sigma}=0$. The
temperature axis is scaled with the nematic-isotropic transition
temperature $T_{NI}$ of the fully occupied model (see
Eq.(\ref{theor18_a})). The dotted line is the spinodal line $T_S$
for micellar demixing, obtained through a high temperature
expansion, Eq.(\ref{theor43}). The dashed line corresponds to the
divergence temperature $T^{*}$ for the nematic correlation length,
Eq.(\ref{theor32}).\label{figura14}}
\end{center}
\end{figure}

\begin{figure}
\begin{center}
\includegraphics[width=0.8\textwidth]{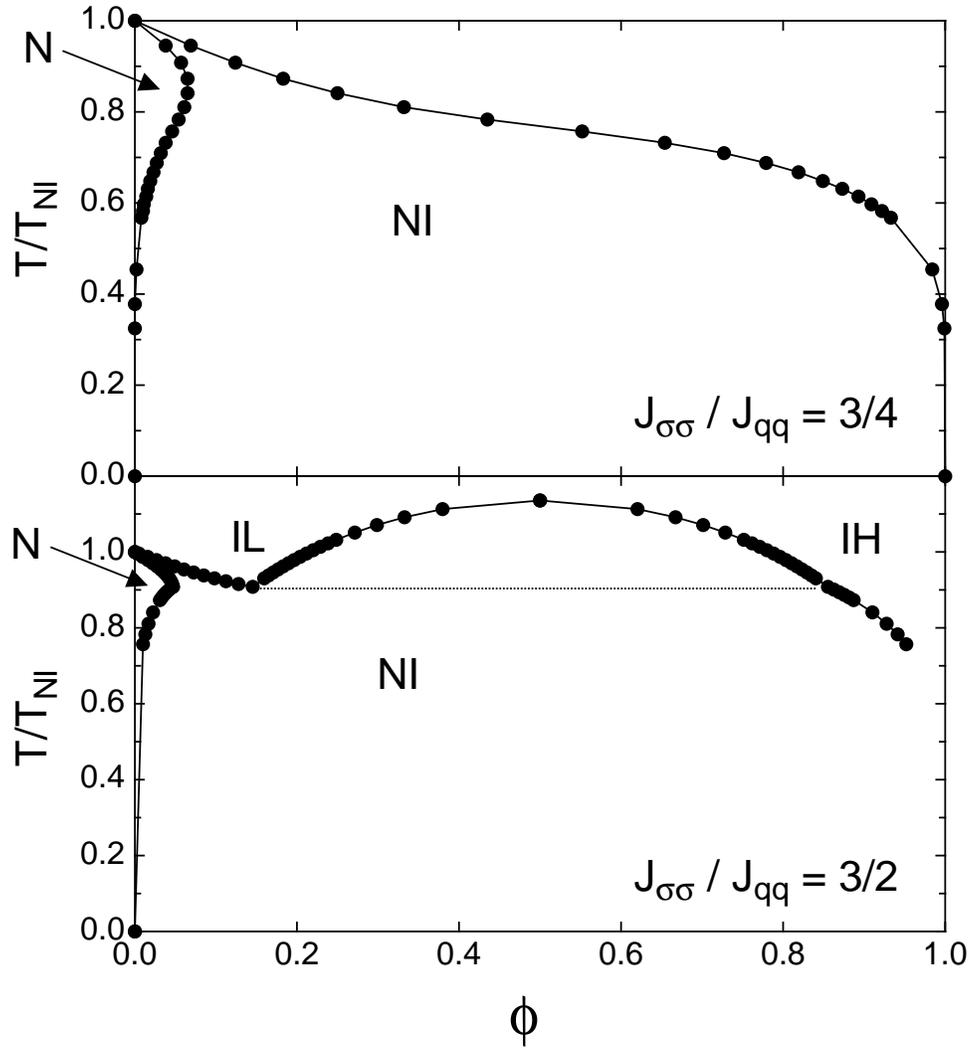}
\caption{Theoretical phase diagram in the case of direct
intermicellar attraction of strength $J_{\sigma
\sigma}/J_{qq}=3/4$ (upper panel) and $J_{\sigma
\sigma}/J_{qq}=3/2$ (lower panel), where the value corresponding
to $\phi=0.5$ is the demixing temperature $T_C$ given by
Eq.(\ref{theor18_b}.) \label{figura15}}
\end{center}
\end{figure}

The final step necessary to obtain the coexistence values $n_N$
and $n_I$ (the nematic and isotropic LC densities at coexistence
respectively) is the Maxwell equal area construction. It is clear
that the resulting phase diagram will be influenced by the
competition between the usual pure nematic-isotropic transition
(associated to the free energy Eq. (\ref{theor14})) and the usual
lattice gas transition (whose free energy is Eq.(\ref{theor16})).
We shall study the isotropic-nematic transition as a function of
the micelle molar fraction $\phi=1-n$ and on the ratio $J_{\sigma
\sigma}/J_{qq}$. The results are depicted in Figs. \ref{figura14},
\ref{figura15} in the case $J_{\sigma \sigma}/J_{qq} = 0$ (no
direct micelle-micelle interaction), $J_{\sigma
\sigma}/J_{qq}=3/4<1$ (weak direct micelle-micelle interaction)
and $J_{\sigma \sigma}/J_{qq}=3/2>1$ (strong direct
micelle-micelle interaction) respectively. The temperature axis is
scaled in terms of the nematic-isotropic (NI) transition
temperature $T_{NI}$ of the fully occupied counterpart
  (FLL), which turns out to be
\begin{equation}
\frac{k_B T_{NI}}{J_{qq}} = 1.9813\ldots \label{theor18_a}
\end{equation}
In Fig. \ref{figura14}, there is no direct coupling among the
micelles ($J_{qq} = 0$) although there is an effective interaction as we shall
see later on. As the micelles concentration $\phi$ increases, the
isotropic phase (I) is depressed until it vanishes at $\phi=1$, a
nematic re-entrant phase (N) appears in the small $\phi$ region,
and the two phases are separated by a nematic-isotropic region
(NI). In figure \ref{figura15} the direct micelle-micelle coupling
is weaker that the LC counterpart and the liquid-gas transition
lies well inside the isotropic-nematic coexistence phase. Hence,
the phase diagram is topologically similar to the previous one. At
sufficiently high values of the direct coupling (Fig.
\ref{figura15} lower panel) the liquid-gas transition
characteristic of the lattice gas model preempts the
isotropic-nematic transition at intermediate micelle molar
fraction. In this case, the consolution curve separating an isotropic
phase with low micellar concentration (IL) from an isotropic
phase with high micellar concentration (IH),
is symmetric because of the vacancy/filled site symmetry,
and its highest point corresponds to the liquid-gas critical
temperature $T_c$
\begin{equation}
\frac{k_B T_C}{ J_{qq}} =\frac{3}{2} \frac{J_{\sigma
\sigma}}{J_{qq}} \label{theor18_b}
\end{equation}

The above phase diagram, as derived from a mean-field theory, is
completely consistent with recent Monte Carlo numerical
simulations on the same model \cite{pre42}, although the numerical
values are not, as one expects from a mean-field-like theory such
as the one treated here.

\subsubsection{Correlation Functions}\label{Correlation Functions}
In order to make a connection with the experimental results
probing both the paranematic fluctuations and the micellar density
fluctuations, in this Section we compute the correlation lengths. This can be
done rather conveniently within our mean field theory by computing
the correlation functions as follows. The micelle-micelle
correlation function is defined as

\begin{eqnarray}
g^{\sigma \sigma}_{x y}&=&\langle(1-\sigma_x)(1-\sigma_y)\rangle -
\langle (1-\sigma_x) \rangle \langle (1-\sigma_y)\rangle \\
&=& \langle \sigma_x \sigma_y\rangle - \langle \sigma_x \rangle
\langle \sigma_y \rangle \label{theor19}
\end{eqnarray}

This can be easily computed in momentum space

\begin{equation}
g^{\sigma \sigma}_{x y}=\frac{1}{N} \sum_\textbf{p}
e^{i\textbf{p}(\textbf{x}-\textbf{y})} \hat{g}^{\sigma \sigma}
(\textbf{p}) \label{theor20}
\end{equation}

as its Fourier transform displays the following very simple form

\begin{equation}
g^{\sigma \sigma} (\textbf{p}) = \frac{1}{n^{-1}
(1-n)^{-1}-2K_{\sigma \sigma} \sum^3_{\mu = 1} \cos{p_{\mu}a}}
\label{theor21}
\end{equation}

The correlation length for the micellar density $\xi_M$ can be
immediately inferred from the behavior of $g^{\sigma \sigma}
(\textbf{p})$ in the continuum limit, leading to

\begin{equation}
g^{\sigma \sigma}_{x,y}=\frac{1}{4 \pi K_{\sigma \sigma}}
\frac{e^{-|x-y|/ \xi_M}}{3 / \xi_M} \label{theor22}
\end{equation}

with

\begin{equation}
\frac{\xi_M}{a}=\sqrt{\frac{K_{\sigma \sigma}/9}{n^{-1}(1-n)^{-1}-
6 K_{\sigma \sigma}}} \label{theor23}
\end{equation}

It is worth noticing that $\xi_M \rightarrow 0$ as  $K_{\sigma
\sigma}\rightarrow 0$, meaning that, at the mean field level,
there is no micelle-micelle
interaction, in the absence of an explicit coupling among them.
Accordingly, the above expression diverges at the (reduced) spinodal
temperature $T_S$
\begin{equation}
\frac{k_B T_{S}}{J_{qq}} = 6 n (1-n) \frac{J_{\sigma
\sigma}}{J_{qq}} \label{theor24}
\end{equation}
which coincides with that of a simple lattice gas model. Only going
beyond the mean field approximation, as described in the next
Section, a LC mediated intermicellar interaction can be described.

It is also interesting to evaluate how the micelle-LC coupling
intrinsically present in this system affects the orientational
pretransitional behavior. The nematic correlation function is
defined as

\begin{equation}
G^{\alpha \beta \mu \nu}_{xy} = \langle \sigma_x Q_x^{\alpha
\beta} \sigma_y Q_y^{\mu \nu} \rangle - \langle \sigma_x
Q_x^{\alpha \beta} \rangle \langle \sigma_y Q_y^{\mu \nu} \rangle
\label{theor25}
\end{equation}

which can be conveniently expressed in terms of the derivative of
the Legendre transform of $F_{MF}$

\begin{equation}
[G^{\alpha \beta \mu \nu}]^{-1}=\frac{\partial^2}{\partial
q^{\alpha \beta}_x \partial q_y^{\mu \nu}} [\beta
\hat{\Gamma}_{MF}] \label{theor26}
\end{equation}

In the frame in which Eq.(\ref{theor7}) holds, it is then easy to
calculate $G_{xy}^{3333}\equiv G_{xy}$. This can be computed along the same lines as for the micellar
case. One obtains, after some algebra, an expression similar to
Eq.(\ref{theor22}) with the orientational correlation length
$\xi_N$ given by

\begin{equation}
\frac{\xi_N}{a}=\sqrt{\frac{3B(K_{qq})}{A(K_{qq},q,n)-3B(K_{qq})}}
\label{theor27}
\end{equation}

where

\begin{equation}
A(K_{qq},q,n)=12 K_{qq} \left\{ 1 - 2 K_{qq} \left[ \frac{9}{16} n
(1-n) \Phi^2 (\frac{27}{2}K_{qq} q) + \frac{81}{16} n R_4
(\frac{27}{2}K_{qq} q) \right] \right\} \label{theor28}
\end{equation}

\begin{equation}
B(K_{qq})= 2 K_{qq} \label{theor29}
\end{equation}

and where we have defined the quantity

\begin{equation}
R_4(\lambda)=\frac{I_4(\lambda)}{I_0(\lambda)} -
\frac{I_2^2(\lambda)}{I_0^2(\lambda)} \label{theor30}
\end{equation}

In particular in the isotropic phase $(q=0)$ this yields

\begin{equation}
\left. \frac{\xi_N}{a}\right|_{q=0} = \sqrt{\frac{1}{1-\frac{9}{5}
    K_{qq} n}}
\label{theor31}
\end{equation}

diverging at the temperature
\begin{equation}
\frac{k_B T^{*}}{ J_{qq}} = \frac{9}{5} n \label{theor32}
\end{equation}
and reducing, in the limit $n \to 1$, to the correct value of the FLL.
The $T^{*}(\phi)$ line depicting the divergence temperature for the
paranematic correlation length is shown in Fig. \ref{figura14}
as a function of the micellar density.

\subsection{High temperature expansion}\label{High temperature expansion}

As previously remarked, one important question arising from the
mean-field calculation above is whether there exists an effective
micelle-micelle interaction, even in the absence of a direct
coupling, which is induced by the fluctuation of  LC degrees of
freedom. One possible indicator of this is the presence of a
micelle spinodal line which signals the approach of a demixing
instability, i.e. the divergence of the correlation length, and we
saw that there is no spinodal line in the absence on an explicit
coupling among the micelles. We attribute this to a deficiency of
the mean field calculation.

To see that this is the case, we proceed as follows. First we
split the original Hamiltonian Eq.(\ref{theor3}) at zero field
($h^{\alpha \beta}=0$) in two parts
\begin{eqnarray}
H\left(\{ Q,\sigma \} \right) &=& H_{LC}  \left(\{ Q,\sigma
\}\right)+ H_{LG}  \left(\{\sigma \}\right) \label{theor33}
\end{eqnarray}
where
\begin{eqnarray}
  -\beta H_{LC}\left(\{ Q,\sigma \}\right)  &=& K_{qq}
\sum_{\alpha \beta} \sum_{\langle xy \rangle} Q_{x}^{\alpha\beta}
Q_{y}^{\beta\alpha} \sigma_x \sigma_y \label{theor34}
\end{eqnarray}
and where the remaining term  $H_{LG}  \left(\{\sigma \}\right)$
is a lattice gas hamiltonian independent on the LC degrees of
freedom $\{ Q \left(\hat{s}\right)\}$. We then introduce an
effective micelle-micelle reduced Hamiltonian $-\beta H_{eff}(\{
\sigma\})$ by integrating out the LC degrees of freedom of the
original Hamiltonian Eq.\ref{theor33}

\begin{eqnarray}
e^{-\beta H_{eff}\left(\{\sigma \}\right)}&=& e^{-\beta
  H_{LG}\left(\{\sigma \}\right)} \frac{\int D \hat{s} \;
e^{-\beta  H_{LC}\left(\{ Q,\sigma \}\right) }}{\int D \hat{s}}
\label{theor35}
\end{eqnarray}

where we have defined

\begin{eqnarray}
\int D \hat{s} &=& \int \prod_x d \hat{s}_x = \prod_x
\int_0^{2\pi} d \phi_x \int_{-1}^{+1}   d(\cos \theta_x)
\label{theor36}
\end{eqnarray}

In the high temperature limit $T >> T_c$ then $K_{qq} << 1$ and
thus
\begin{eqnarray}
&&\int D \hat{s} \; \; e^{K_{qq} \sum_{\alpha \beta} \sum_{\langle
xy \rangle} Q_{x}^{\alpha\beta} Q_{y}^{\beta\alpha} \sigma_x
\sigma_y } = \\ \nonumber &&\int D \hat{s}  \left[1+ K_{qq}
\sum_{\alpha \beta} \sum_{\langle xy \rangle} Q_{x}^{\alpha\beta}
Q_{y}^{\beta\alpha} \sigma_x \sigma_y  +
 \frac{1}{2} K_{qq}^2 \sum_{\alpha \beta} \sum_{\langle xy \rangle}
Q_{x}^{\alpha\beta} Q_{y}^{\beta\alpha} \sigma_x \sigma_y
\sum_{\mu \nu} \sum_{\langle z w \rangle} Q_{z}^{\mu \nu}
Q_{w}^{\nu \mu} \sigma_z \sigma_w + \ldots \right] \label{theor37}
\end{eqnarray}

Using the following results

\begin{equation}
\int d \hat{s} Q^{\alpha \beta} (\hat{s}) = 0 \label{theor38}
\end{equation}

\begin{equation}
\int d \hat{s} Q^{\alpha \beta} (\hat{s}) Q^{\mu \nu} (\hat{s})= -
\frac{2}{5} \pi \delta^{\alpha \mu} \delta^{\mu \nu} + \frac{3}{5}
\pi (\delta^{\alpha \mu} \delta^{\beta \nu} + \delta^{\alpha \nu}
\delta^{\beta \mu}) \label{theor39}
\end{equation}
we obtain, to first order in $K_{qq}$ and after rexponentiating the
result, a new lattice gas Hamiltonian
\begin{equation}
- \beta H_{eff} \left(\{\sigma \}\right) = K_{\sigma \sigma}^{eff}
\sum_{\langle xy \rangle} \sigma_x \sigma_y + \sum_x \mu^* (1-
\sigma_x) \label{theor40}
\end{equation}
where
\begin{equation}
K^{eff}_{\sigma \sigma} = K_{\sigma \sigma} + \frac{9}{40}
K^2_{qq} \label{theor41}
\end{equation}
which amounts to an effective coupling
\begin{equation}
J^{eff}_{\sigma \sigma} = J_{\sigma \sigma} + \frac{9}{40}
\frac{J^2_{q q}}{k_B T} \label{theor42}
\end{equation}
Notice that $J^{eff}_{\sigma \sigma} $ depends upon the temperature $T$.

In the case $K_{\sigma \sigma}=0$, we can calculate the spinodal
line $T_{S}(n)$ on the basis of the mean field solution of the
standard lattice gas model. We obtain
\begin{equation}
\frac{k_B T_S}{J_{qq}}=\sqrt{\frac{27}{20}n(1-n)} \label{theor43}
\end{equation}

which is also depicted in Fig.  \ref{figura14}. Note that this
result differes from that of a simple lattice gas
(Eq. \ref{theor24}) because of the $T$ dependence of the effective
coupling.

This analysis shows effective micelle-micelle interactions, can
be computed by eliminating the LC degrees of freedom from
the problem, as done here through a calculation correct to
a first order in a high temperature expansion.

\subsection{Radial coupling}\label{Radial coupling}
Before ending this section, we wish to comment on a possible
additional term appearing in the general Hamiltonial
(\ref{theor3}). As described in Ref.\cite{pre6}, the properties of the present
system can be strongly influenced by a directional coupling
between the LC and the micelles. Hence one is tempted to include a
tendency of the LC to be oriented along the radial direction of
the micelles whenever the LC lie in their proximity, as a possible
consequence of the tendency for homeotropic anchoring of DDAB.

To account for this we add the following additional term $-\beta
\Delta H_A$ to the Hamiltonian in Eq.(\ref{theor3})

\begin{equation}
- \beta \Delta H_A = \frac{K_{q \sigma}}{a{^2}} \sum_{\alpha
\beta} \sum_{\langle xy \rangle} [(1- \sigma_x) \sigma_y
Q_y^{\beta \alpha} (\textbf{y} -
\textbf{x})^{\alpha} (\textbf{y} - \textbf{x})^{\beta} \\
 + \sigma_x (1- \sigma_y) Q_x^{\alpha
\beta} (\textbf{x} - \textbf{y})^{\alpha}
(\textbf{x}-\textbf{y})^{\beta} ] \label{theor44}
\end{equation}

\begin{figure}
\begin{center}
\includegraphics[width=0.8\textwidth]{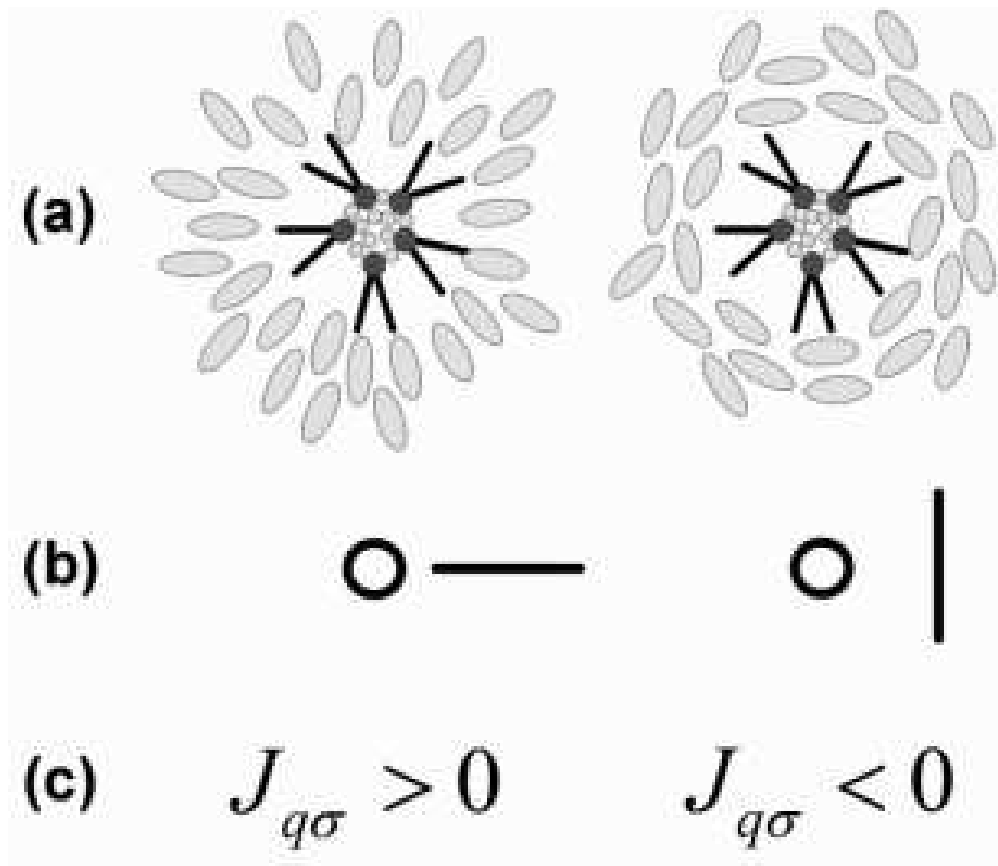}
\caption{Possible Liquid crystal couplings at the micelle surface.
Left: radial coupling. Right: tangential coupling. a: 2D pictorial
view of the couplings. b: schematization of the coupling as in the
lattice spin model, where circle represents an empty site and the
segments the headless spins. c: coefficient $K_{q\sigma}$ of eq.
\ref{theor44} expressing the coupling strenght.\label{figura16}}
\end{center}
\end{figure}

As  described in Fig.  \ref{figura16}, the sign of
$K_{q\sigma} = \beta J_{q \sigma}$ identifies
the preferred orientation of the LC with respect to its distance
from the center of the neighboring the micelle.
If $J_{q \sigma}>0$ the LC at the LC-micelles
interface tend to align radially. It
is easy to see that, at the mean field level,  the above term does
not affect the phase diagram.
This is because mean field theory effectively eliminates the site
dependence of the order parameters.

Things are different when the same coupling is investigated through
high temperature expansion. In this case, the radial coupling leads
to non-trivial three-body terms. This, at variance with the
mean-field theory, indicates the possible relevance of the LC
anchoring on the micellar surface.

\section{Discussion}\label{discussion}

\subsection{Comparison between experimental results and theoretical
model}\label{Comparison between experimental results and
theoretical model}

The existence of Casimir-style, fluctuation mediated
inter-micellar attractions, demonstrated by the high temperature
expansion in the case of no direct coupling ($K_{\sigma
\sigma}=0$), is a key result that enables us to interpret the
observation of pretransitional micellar density fluctuations in
systems where paranematic fluctuations are also observed. Indeed,
the theoretical modelling enables us to understand the
inter-micellar attraction documented in the experiments as
interactions mediated by paranematic fluctuations.

More specifically, we argue that the experimental conditions we
investigated correspond to $K_{\sigma \sigma} \approx 0$ and $K_{q
\sigma} \approx 0$. The simultaneous presence of micellar and
orientational fluctuations, and the fact that upper critical
solution points have never been observed in w/o microemulsions,
make us reject as unlikely (but not impossible) that a direct
interaction may be at the origin of the observed behavior. Such an
interaction should in fact be independent from the ordering of the
solvent and of a different nature with respect to those leading to
lower critical consolution point in other systems \cite{pre31}. On
the other hand, the fact that the single molecule dynamics
detected by EBS is unaffected by the presence of the micelles (see
Section \ref{me EBS}), together with the nanosize of the droplet
radius, way shorter than typical extrapolation lengths for surface
anchoring, strongly support the notion that the 5CB anchoring at
the micellar surface is very weak.

Therefore, we read the observed increase in $\langle I_M \rangle$
and decrease in $D_M/D_0$ on the basis of the theoretical
predictions that the softening of the micellar system (increase of
micellar osmotic compressibility and associated slowing down) is
expected on the only basis of the micellar dilution effect of the
local nematic matrix. The existence of a bell-shaped spinodal
curve in Fig. \ref{figura14}, suggests that a similar curve could
be present in the experimental phase diagram as well (Fig.
\ref{figura2}), although hidden into the coexistence region. Thus,
on lowering T, i.e. on approaching the spinodal line increasing
micellar density fluctuation are expected. Accordingly, we have
fitted $\langle I_M \rangle$ vs. $T$ with $(T- T_S)^{-1}$. The
extracted values for $T_S(\phi)$, which suffer of a rather large
indetermination because of the intrinsically limited range of the
fit, are shown in Fig. \ref{figura2}. The two data point appear to
decrease with increasing $\phi$ for $\phi
> 0.075$, but for small experimentally inaccessible $\phi$,  $T_S(\phi)$ must
take a positive slope since the system of highly diluted micelles
necessarily approaches gas-like non-interacting behavior.

The comparison between the experimental phase diagram (Fig.
\ref{figura2}) and the theoretical one (Fig.\ref{figura14})
enlightens a remarkable similarity in all the qualitative
features: re-entrant nematic phase, shape of the coexistence
curve, non-trivial divergence lines determining the orientational
and micellar pretransitional fluctuations. However, no
quantitative comparison can be drawn between the two phase
diagrams because of the approximations at the base of the model
and in the techniques used to solve it. Both the temperature and
the micellar density axes cannot be easily mapped on the
corresponding experimental quantities. For example, the
$T_{NI}-T^*$ gap for the pure nematic as well as the value of
$\phi$ corresponding to the maximum in the spinodal curve are both
much larger than the corresponding observed quantities. These
shortcomings are consequences, respectively, of the intrinsic
limitations of the mean field approximation  and of having assumed
that micelles and LC molecules have equal volumes .

According to the interpretation here proposed, the expulsion of
micelles by orientational ordering, explicitly manifested at low T
in the macroscopic phase separation, is at the root of the whole
phase behavior of $5DH \mu em$. Nematic correlations appear more
easily where the local micelle concentration is lower, and in turn
force out micelles, thus further reducing the local amount of
annealed disorder and increasing the local $T^*$, with the
consequent increase of local nematic order. Theoretical
considerations indicate that sufficiently small impurities
incorporated in the nematic phase display attractive long ranged
interactions due to orientational fluctuations of the LC
\cite{pre50}. This is a consequence of the impurity induced
depression of the local nematic order which are thus segregated to
minimize the total free energy, a mechanism that bears some
resemblance to the prewetting induced colloidal aggregation
\cite{pre51}, or to the depletion forces in critical systems
\cite{pre52}. We argue that the same mechanism should be present
in the isotropic phase but with an exponential cut-off at $\xi_N$.
This notion is corroborated by the experimental coincidence
between the average intermicellar distance d and the maximum value
of $\xi_{N,max} = \xi_0 [T^*/(T_{DM}-T^*)]^{\frac{1}{2}}$, with
$\xi_0 \sim 10 \Ang$ : for $\phi = 0.075$, $\xi_{N,max} \sim 95
\Ang$, $d \sim 58 \Ang$; for $\phi = 0.15$, $\xi_{N,max} \sim 75
\Ang$, $d \sim 74 \Ang$).

\subsection{Concentration dependence of the transition
temperatures}\label{Concentration dependence of the transition}

All the observed $5DH \mu em$ behavior appears determined by three
relevant temperatures: the demixing temperature $T_{DM}(\phi)$,
the divergence temperature for paranematic fluctuations
$T^*(\phi)$, the spinodal temperature $T_S(\phi)$. We want here
discuss the origin of their $\phi$ dependence.

One of the most striking features of the phase diagram of the $5DH
\mu em$ is the high negative slope of $T_{DM}(\phi)$ and
$T^*(\phi)$: $T_{NI} -T_{DM}(\phi) \sim 4.4 K$ and $T_{NI} -
T^*(\phi) \sim 7 K$ for the $5DH \mu em$ $\phi=7.5 \%$; $T_{NI} -
T_{DM}(\phi) \sim 7.6 K$ and $T_{NI} - T^*(\phi)\sim 11 K$ for the
$5DH \mu em$ $\phi=15 \%$. Thus $dT_{DM}(\phi)/d \phi \sim - 60 K$
and $dT^*(\phi)/d\phi \sim - 80 K$. These numbers are quite larger
than the T shift in the position of the specific heat $(c_p)$ peak
observed by studying LC with nanosized quenched disorder:
$dT_{NI}(\phi)/d \phi \sim - 7.1 K$ for 8CB embedded in silica
aerogel \cite{pre36} and $dT_{NI}(\phi)/d\phi \sim - 8.2 K$ for
$8CB$ incorporating soft gels of aerosil particles \cite{pre37}.
The origin of such a difference will also be discussed in the
following.

\subsubsection{Demixing temperature $(T_{DM})$}\label{Demixing temperature}

The determination of the presence and size of the droplets enables us
to better discuss the low concentration part of the phase
diagram. Indeed, the value of $\beta_I$, too low if calculated on
the basis of the molar fraction of monomeric DDAB (see Section III.A), takes a much
higher value when referred to the molar fraction of micellar
aggregates, which can be calculated from the mass of a single
micelle. The
radius that micelles would have if DDAB were compacted in a sphere
was estimated to be about $13.5 \Ang$ (see Sec. III.C.3).
Based on this value, the slopes of the isotropic and nematic phase
boundary vs. contaminant fraction are, respectively $\beta_I \sim
-4.17$ and $\beta_N \sim -31.85$. We thus obtain $(\beta_N^{-1} -
\beta_I^{-1}) \sim 0.21$, in a much better agreement with the
prediction of $( \beta_N^{-1} - \beta_I^{-1}) \sim 0.25$. The
numerical difference between the two numbers indicates that the
calculation partially underestimates the slope of the isotropic
phase boundary.

The same relationship between phase boundary and contaminant
concentration can be reformulated under a different prospective,
which offers a better insight on its physical basis. By
approximating $\beta_N^{-1} \sim 0$, the occurrence of the $5DH
\mu em$ demixing transition coincides with a transformation of a
fraction of the LC from isotropic to nematic and, at the same
time, with the segregation of the inverted micelles in the
isotropic phase. By balancing the gain in free energy density for
the nematic ordering of 5CB $( \Delta G_{NI}(T))$ against the work
done against the osmotic pressure $\Pi$ of the micelles, we obtain

\begin{equation}
\Delta G_{NI}(T)=\Pi(\phi) \label{}
\end{equation}

$\Delta G_{NI}(T)$ can be computed on the basis of the
Landau-deGennes expansion, whose coefficients for bulk 5CB have
been determined \cite{pre53}. Indeed, since the micelle-disordered
isotropic cannot differ much from the unperturbed isotropic
(especially because surface coupling is very weak), and since the
nematic phase in the $5DH \mu em$ is almost micelle-free,  $\Delta
G_{NI}(T)$ calculated for bulk 5CB is a good approximation for the
difference in free energy between nematic micelle poor phase and
isotropic micelle rich one. The osmotic pressure, on the other
side, can be calculated on the basis of $\phi$ and R, the micellar
radius, and by assuming a perfect gas behavior. With this with
these ingredients, it is possible to extract $\beta_I$ in the
limit of small $\phi$ . With this crude estimate we obtain $\beta_I \sim -6.5$.
Although overestimating the measured $\beta_I$, this figure is
close enough to the data to indicate that indeed the temperature depression
of the demixing transition can be understood as a direct
consequence of the very nature of the demixing. This fact once again
confirms the appropriateness of describing the $5DH \mu em$ as a
pseudo-single component system.

\subsubsection{Paranematic divergence temperature $(T^*)$}\label{Paranematic divergence temperature}
According to the model $T^*(\phi) = T^*(\phi = 0) \cdot (1 - \phi
)$ (see Section \ref{Theoretical model}). The decrease of $T^*$
with $\phi$ can be easily interpreted as a dilution effect on the
coupling strength. Indeed, the dilution brings about a decrement
in the average coordination number of each spin. As a consequence
$\overline{J_{qq}}$, the coupling coefficient averaged on all
bonds, also decreases. Since $\phi$ is the probability to find a
micelles as nearest neighbor of each given spin,
$\overline{J_{qq}}=J_{qq} \cdot (1-\phi)$. Hence, it appears that
the easiest interpretation of the observed $T^*(\phi)$ is to picture
it as a direct consequence of a dilution induced weakening of the
molecular field, which, in the experimental case, is associated to
the presence of interfaces reducing the coordination number of the
5CB molecules. A possible estimate of the fraction of 5CB
molecules at contact with the micellar surface is given by the
volume between $13.5 \Ang$ and $20 \Ang$ from the center of the
micelle, i.e. by the difference between the volume
hydrodynamically attached to the micelle and the total water-DDAB
volume. This would imply that a fraction of about $f = 3.2 \cdot
\phi$ of the total 5CB molecules experience reduced coordination
because of the proximity of DDAB molecules. Comparing $T^*(\phi) =
T^*(\phi = 0) \times (1 - \phi)$ with the experimentally observed
$dT^*(\phi) / d \phi  \sim - 80 K$, suggests that the coordination
number of 5CB molecules at the DDAB interface is reduced of 8 \%.
This number appears reasonable, since we expect it to be lower
than the simplest available analogy of a hard sphere crystal
limited by a plane. In that case spheres facing the plane have a
coordination of 9 while in the bulk the coordination is 12,
yielding a decrease of $25 \%$ of the coordination number. This
simple evaluation, despite its extreme roughness, supports the
notion that the concentration dependence of $T^*$, and thus the
whole T downshift in the pretransitional paranematic behavior, is
due to dilution induced weakening of the intermolecular coupling.

\subsubsection{Comparison with LC in silica gels}\label{Comparison with LC in silica gels}
The N-I transition in LC with dispersed silica nanostructures,
either aerogel or aerosil, displays a specific heat $c_p$ peak
downshifted in T with respect to the corresponding
transition in the bulk and downshifted in T. In those systems
it is found that the whole paranematic behavior is unmodified but "rigidely"
shifted in temperature (\cite{pre36}, \cite{pre37}). It is quite remarkable that
such observed $dT^*/d \phi$ in the aerogel/aerosil LC systems
is about ten times smaller than in $5DH \mu em$. The different nature of the
disordering provided by the two structures hosted in the LC,
mobile "annealed" disorder in $5DH \mu em$ and static "quenched"
disorder in nanosilica-LC materials, cannot be the origin of this
large effect. This is because the different degrees of freedom are
highly mobile in the high temperature phase, thus preventing a
clear distinction between a frozen impurity and an equilibrated
one.

The difference between the two systems can, at least in part, be
understood on the basis of different dilution induced weakening of
intermolecular coupling, along the line of the discussion in the
previous section. X-ray characterization of the empty aerogel
structure - not available in the aerosil case - enable extracting
the specific interface extension of the structures, expressed in
the S/V ratio, where S is the silica surface area and V the
aerogel volume \cite{pre54}. For a particular family of silica
aerogel extensively characterized in the past, it is found that,
approximately, $S/V  = 0.11 \Ang^{-1}$ \cite{pre38}. It follows
that the fraction of LC molecules at the silica interface is
approximately $f \sim S/V \cdot \sqrt[3]{v_{LC}} \sim 0.8$ , where
$v_{LC}$ is the volume of a single LC molecule. Thus, at equal
volume fraction, the silica aerogel is characterized by an
interface about three times smaller than the $5DH \mu em$, which
would imply a value of $dT^*/d \phi$ three times smaller, thus
only in part accounting for the experimentally observed
difference. This fact indicates that some other factor plays a
relevant role in determining the $T^*$ shift, possibly related
either to the different surface coupling of the two surfaces or to
the different geometry of the disordering hosts: well dispersed in
the $5DH \mu em$, while necessarily interconnected in the aerogel.

\section{Conclusions}\label{Conclusions}

In this article we have presented a systematic investigation of a
system where thermodynamically stable nanoparticles are dispersed
in a thermotropic liquid crystal. As it appears from the
experiments, the self aggregated nature of the particles does not
play any observable role, in agreement with the notion that such
self assembly is governed by forces larger than the molecular
field strength provided by nematic ordering. This system enables
studying the effect of annealed disorder on the phase diagram of
the liquid crystal, and, at the same time, thanks to the large
optical contrast provided by the droplets' water core, it also
allows direct observation of the statistical properties of the
dispersed nanoparticles.

We found that the system exhibits a phase transition combining the
demixing of the nano-droplets with nematic ordering in one of the
coexisting phases. The phase diagram also features an intriguing
reentrant nematic phase. This feature indicates a temperature
dependent solubility of nano-droplets in the nematic phase.

The characteristic frequency of the electric birefringence spectra
expressing the time to redirect the molecular axis, shows that the
dispersed nanoparticles do not alter appreciably the local
molecular environment which remains bulk-like. This fact supports
the notion that, mostly because of their nano-size, microemulsion
droplets do not provide significant anchoring on the neighboring
LC molecules. Hence, the observed effects of the nanoinpurities on
the IN phase transition have to be understood simply as
consequences of dilution, in strong analogy with the most
conventional understanding of the effects of molecular
contaminants. Along this line we have here presented a theoretical
analysis aimed to determine the phase behavior of a diluted
Lebwohl-Lasher spin lattice model, where LC molecules are mimicked
by pointless spin vectors on a cubic lattice and micelles are
vacancies on the same lattice. The spin variables interact via a
Maier-Saupe orientational coupling, provided that nearest
neighbour pairs are occupied by LC variables. This system has been
solved by mean-field theory and high-temperature expansion.

Experimental observations and theoretical predictions nicely match
offering a simple ground to interpret all the basic features of
the system.

The rather complex topology of the observed phase diagram
(demixing transition and re-entrant nematic phase) simply follows
from the tendency of liquid crystal towards orientational order
with no need for either micelle-micelle or micelle-LC direct
interactions. The dependence of phase boundaries and divergence
temperatures on the concentration of impurities conveys the mutual
effects of orientational order and nanoparticle dispersion.
Furthermore, the study of the light scattered by the
nanodroplets in the isotropic phase close to the $IN$ transition,
enlightens the presence of intermicellar attractive forces which
the theoretical model enables interpreting as a consequence of the
paranematic orientational fluctuations. This can be seen by
eliminating the LC variables in the Hamiltonian of the spin model,
approximated to the leading order in a high temperature expansion.
An effective temperature dependent micelle-micelle interaction is
found, whose strength depends on the orientational coupling
between the LC molecules. This fluctuation mediated interaction is
intrinsic to the very presence of nanoimpurities, necessarily
diluting the LC, while the addition of possible surface anchoring
does not change the picture within a mean-field theory, but has a
non-trivial effect within our high-temperature expansion.
Accordingly, it appears interesting to explore how the phase
diagram is modified by increasing the droplet size and the
hydrophobic portion of the surfactants, which should lead to
progressively stronger anchoring.

We acknowledge support from NATO (Collaborative Linkage Grant) and
from MIUR (PRIN 2004024508)

\clearpage

\clearpage

\clearpage

\clearpage

\clearpage

\clearpage

\clearpage

\clearpage

\clearpage

\clearpage

\clearpage

\clearpage

\clearpage

\clearpage

\clearpage

\clearpage

\clearpage

\end{document}